\documentclass[twocolumn, twocolappendix]{aastex63}

\usepackage{etoolbox}
\usepackage{graphicx}	
\usepackage{multirow}
\usepackage{cleveref}
\usepackage{CJK}
\usepackage{stfloats}

\newcommand{\kms}{km s$^{-1}$}
\newcommand{\cmN}{cm$^{-2}$}

\newcommand{\lam}{$\lambda$}

\newcommand{\lya}{\mbox{Ly$\alpha$}}

\newcommand{\civ}{\mbox{C\,{\sc iv}}}

\newcommand{\cii}{\mbox{C\,{\sc ii}}}
\newcommand{\siiv}{\mbox{Si\,{\sc iv}}}

\newcommand{\siii}{\mbox{Si\,{\sc ii}}}
\newcommand{\nv}{\mbox{N\,{\sc v}}}

\newcommand{\ovi}{\mbox{O\,{\sc vi}}}

\newcommand{\oi}{\mbox{O\,{\sc i}}}

\newcommand{\feii}{\mbox{Fe\,{\sc ii}}}

\newcommand{\pv}{\mbox{P\,{\sc v}}}

\received{}
\revised{}
\accepted{}

\submitjournal{ApJ}

\shorttitle{A Catalog of \civ\ Mini-BALs}
\shortauthors{Chen et al.}

\begin{document}

\title{A Catalog of High-Velocity \civ\ Mini-BALs in the VLT-UVES and Keck-HIRES Archives}

\correspondingauthor{Bo Ma}
\email{mabo8@mail.sysu.edu.cn}

\author{Chen Chen}
\affiliation{School of Physics $\&$ Astronomy\\
Sun Yat-Sen University\\ 
Zhuhai 519000, China}
\affiliation{Department of Physics $\&$ Astronomy\\
University of California\\ 
Riverside, CA 92521, USA}

\author{Fred Hamann}
\affiliation{Department of Physics $\&$ Astronomy\\
University of California\\ 
Riverside, CA 92521, USA}

\author{Bo Ma}
\affiliation{School of Physics $\&$ Astronomy\\
Sun Yat-Sen University\\ 
Zhuhai 519000, China}

\author{Michael Murphy}
\affiliation{Centre for Astrophysics and Supercomputing\\
Swinburne University of Technology\\
Hawthorn, Victoria 3122, Australia}

\begin{abstract}

We present a catalog of high-velocity \civ\ \lam 1548,1551 mini-Broad Absorption Lines (mini-BALs) in the archives of the VLT-UVES and Keck-HIRES spectrographs. We identify high-velocity \civ\ mini-BALs based on smooth rounded BAL-like profiles with velocity blueshifts $<$ $-$4000 \kms\ and widths in the range 70 $\lesssim$ FWHM(1548) $\lesssim$ 2000 \kms\ (for $\lambda$1548 alone). We find 105 mini-BALs in 44 quasars from a total sample of 638 quasars. The fraction of quasars with at least one mini-BAL meeting our criteria is roughly $\sim$9\% after correcting for incomplete velocity coverage. However, the numbers of systems rise sharply at lower velocities and narrower FWHMs, suggesting that many outflow lines are missed by our study. All of the systems are highly ionized based on the strong presence of \nv\ and \ovi\ and/or the absence of \siii\ and \cii\ when within the wavelength coverage. Two of the mini-BAL systems in our catalog, plus three others at smaller velocity shifts, have \pv\ $\lambda$1118,1128 absorption indicating highly saturated \civ\ absorption and total hydrogen column densities $\gtrsim 10^{22}$ \cmN. Most of the mini-BALs are confirmed to have optical depths $\gtrsim$1 with partial covering of the quasar continuum source. The covering fractions are as small as 0.06 in \civ\ and 0.03 in \siiv , corresponding to outflow absorbing structures $<0.002$ pc across. When multiple lines are measured, the lines of less abundant ions tend to have narrower profiles and smaller covering fractions indicative of inhomogeneous absorbers where higher column densities occur in smaller clumps. This picture might extend to BAL outflows if the broader and generally deeper BALs form in either the largest clumps or collections of many mini-BAL-like clumps that blend together in observed quasar spectra.

\end{abstract}


\section{Introduction}

Outflows from quasar accretion disks might combine with starburst-driven winds to provide important feedback that regulates star formation and mass assembly in the host galaxies \citep{Silk98, Kauffmann00, King03, Scannapieco04, DiMatteo05, Hopkins08, Ostriker10, Debuhr12, Rupke13, Rupke17, Cicone14, Weinberger17}. The kinetic power needed from quasar outflows to produce important feedback effects on their own are believed to be just $\gtrsim$0.5 per cent of the quasar bolometric luminosities \citep{Hopkins10}, up to $\gtrsim$5 percent of bolometric for a full blowout of the interstellar gas \citep{DiMatteo05, Scannapieco04, Prochaska09}.

Quasar outflows are often studied via blueshifted broad absorption lines (BALs) in the rest-frame UV spectra of quasars. The \civ\ \lam 1548,1551 doublet lines are the strongest outflow lines readily accessible outside of the \lya\ forest. BALs are often defined by \civ\ velocity widths $\gtrsim2000$ \kms\ at flow speeds from a few thousand up to a several tens of thousands of \kms\ \citep{Anderson87, Weymann91, Reichard03, Trump06}. However, quasar outflows produce a wide variety of absorption lines of different velocity widths. These include narrow absorption lines (NALs) with widths that are nominally less than a few hundred \kms, and so-called ``mini-BALs'' that have widths intermediate between BALs and NALs \citep[e.g.,][]{Hamann97, Hamann04, Hamann11, Misawa07, Misawa14}. Quasar spectroscopic surveys have shown that mini-BALs and outflow NALs (as opposed to NALs that form in intervening gas unrelated to the quasars) are substantially more common than BALs in quasar spectra. In particular, roughly 50\% of bright quasars in the Sloan Digital Sky Survey \citep[SDSS,][]{Schneider10} exhibit a \civ\ NAL or mini-BAL outflow absorption line while only $\sim$15\% to $\sim$20\% contain a classic \civ\ BAL \citep[][and refs. therein]{Trump06, Paola08, Nestor08, Wild08, Misawa07, Ganguly08, Knigge08, Hamann12, Paris17, Guo19}. 


Mini-BALs are the least studied of these outflow features. They appear across same range of velocity shifts as BALs, nominally from near 0 to $\sim$30,000 \kms\ form, with prominent absorption in higher ions such as \civ, \nv\ \lam 1239,1243, \ovi\ \lam 1032,1038 \citep{Paola08, Leighly09, Hamann11, Hall11, Paola12, Horiuchi16, Rogerson16, Moravec17, Hamann19a}. However, important questions remain about the physical properties of mini-BAL outflows and their relationships to the outflows measured via BALs and NALs. A simple unified picture might attribute all three line types to a single quasar outflow phenomenon viewed at different angles. For example, if BALs form in the main outflow concentrated near the accretion-disk plane, mini-BALs might appear along the ragged upper edge of these outflows at higher latitudes above the disk \citep{Ganguly01, Hamann04, Chartas09, Elvis12, Hamann12, Matthews16}. However, it is known quasar outflow lines can form across a wide range of distances from the quasars, e.g., at pc to tens of kpc \citep[e.g.,][]{Hamann01, deKool01, Dunn10, Chamberlain15}, and there are known time-dependent effects \citep{Hamann08, Leighly09, Hall11, FilizAk13, Paola13, FilizAk14, Rogerson16, Moravec17} where, for example, broad and deep BAL features can decline over time to become weaker mini-BALs \citep{Hamann04}. 

A more complete understanding of quasar accretion-disk outflows will require larger mini-BAL studies comparable to the survey work already done on BALs and NALs . One obstacle to mini-BAL studies is that the intermediate widths and often shallow depths of the lines make them difficult to detect (and correctly identify) in spectroscopic surveys using moderate resolution and moderate-to-low signal-to-noise ratios, e.g., as in SDSS-I/II and the Baryon Oscillation Spectroscopic Survey \citep[BOSS,][]{Dawson13, Paris17} in SDSS-III \citep{Eisenstein11}. Searches for mini-BALs in these datasets using the \civ\ `absorption index' \citep[AI,][]{Hall02, Trump06, Paris17} produce samples severely contaminated by unrelated intervening absorption lines (e.g., damped \lya\ systems or blends of multiple NALs that have nothing to do with quasar outflows) unless high minimum values on AI are set to avoid this contamination \citep[e.g.,][]{Paola08, Ganguly08, Hamann19a, Hamann19b}. However, large AI values exclude interesting narrow mini-BALs near the ambiguous boundary with NALs, e.g., with velocity widths less than a few hundred \kms . Studies of small quasars samples using higher-resolution echelle spectra confirm that quasar outflows produce a continuous range of line widths from mini-BALs to NALs, while the survey statistics mentioned above suggest that narrower outflow lines are more common than broad ones \citep[e.g.,][]{Hamann97, Hamann97b, Arav01b, Gabel05b, Simon10, Misawa10, Hamann11}.

Another obstacle for studies of broad outflow absorption lines (all BALs and some mini-BALs) is that doublet lines like \civ\ \lam 1548,1551 (with separation 498 \kms ) can blend together so their strength ratios cannot be measured for optical depth constraints. This is problematic because partial covering of the background light source appears to be common in quasar outflows, leading to observed line strengths that depend on the line-of-sight covering fractions in addition to the line optical depths and column densities \citep[Sections 3 and 5 below,][]{Hamann97, Arav05}. In particular, highly saturated lines produced in high-column density outflows can have shallow observed troughs (not reaching zero intensity) in quasar spectra, leading to large uncertainties (generally underestimates) of key outflow properties like the total column densities, mass loss rates, and kinetic energy yields \citep[see also][for recent discussions]{Moravec17, Hamann19a, Hamann19b}. 

Severe line saturation and large total column densities can sometimes be identified via measurements of lines from low-abundance ions like \pv\ \lam 1118, 1128 \citep{Hamann98, Hamann03, Leighly09, Chamberlain15, Capellupo14, Capellupo17, Moravec17, Hamann19a}. Photoionization models indicate that the \pv\ and \civ\ ions coexist spatially in quasar outflows, but the \pv\ line optical depths are roughly $\sim$1000 times smaller than \civ\ due to the lower abundance (assuming roughly solar abundance ratios and ionization by a standard quasar spectrum). Measurements of \pv\ absorption lines therefore indicate that strong transitions like \civ, \siiv , \ovi, and others are highly saturated and that the total column densities in the outflows are conservatively $\log N_H(\textrm{cm}^{-2}) \gtrsim 10^{22}$ \citep{Hamann98, Leighly11, Borguet13, Chamberlain15}. The recent study by \citep{Hamann19a} using median composite spectra of quasars in the BOSS survey shows that \pv\ absorption is common in all types of weak and strong BAL outflows and that it is weaker but still present in typical mini-BAL systems. There are also numerous measurements of \pv\ absorption in individual BAL quasars \citep[e.g.,][]{Capellupo17}. However, \pv\ is much more difficult to detect in weaker and narrower mini-BAL systems due to blending problems in the \lya\ forest. There is, to our knowledge, only one reported case of \pv\ absorption in a mini-BAL outflow \citep[in the low-redshift quasar PG1411+442,][]{Hamann19b}. 

These difficulties provide further motivation to find and study narrow outflow lines where the common doublets like \civ\ are resolved. In this paper, we present a catalog of \civ\ mini-BALs and other narrow outflow lines in archival high-resolution and high signal-to-noise ratio spectra of 638 quasars measured with the UV-Visual Echelle Spectrograph (UVES) on the Very Large Telescope (VLT) or the High Resolution Echelle Spectrometer (HIRES) on the W. M. Keck Observatory telescope. This unique dataset allows us to detect and measure \civ\ mini-BALs that are considerably weaker and narrower than any previous outflow line survey. We deliberately cross the ambiguous boundary from mini-BALs to NALs to include in our catalog all \civ\ lines narrower than BALs that can be readily attributed to quasar outflows based on their smooth rounded BAL-like profiles. Many of the lines included would be considered NALs or ``outflow NALs" in previous work. However, for convenience and because of their related origins in quasar outflows, we will refer hereafter to all of the \civ\ outflow lines in our catalog as mini-BALs. We also report on other lines detected in the same outflow systems, notably \siiv, \nv, \ovi, and \pv, which provide important constraints of the outflow ionizations, covering fractions, and column densities. 

Section 2 below describes the UVES and HIRES datasets and our mini-BAL selection criteria. Section 3 describes fits to the \civ\ mini-BAL profiles that yield measurements of the line full widths at half maximum (FWHMs), rest equivalent widths (REWs), and covering fractions if available. Section 4 presents the basic results and statistical analysis. Section 5 provides a brief summary and discusses the broader implications of our study for nature of quasar outflows and their possible role in feedback to galaxy evolution. Throughout this paper, we adopt a cosmology with $H_0=71$ \kms\ Mpc $^{-1}$, $\Omega_M=0.27$ and $\Omega_{\Lambda}=0.73$.

\section{Datasets \& mini-BAL selection}

\subsection{Quasar Spectra}

The parent sample of quasars for our study comes from two datasets of archival high-resolution spectra obtained with VLT-UVES \citep{Murphy19} and Keck-HIRES \citep{OMeara15}. The VLT-UVES spectra from \citet{Murphy19} are all 468 quasars (468 spectra) first observed with UVES before mid-2008. All exposures of those 468 quasars recorded before late 2016 were combined to form the final UVES spectra. The Keck-HIRES spectra from \citet{OMeara15} are the first data release of the KODIAQ survey containing all 170 quasars (247 spectra, some quasars have multiple spectra obtained at different times) observed with HIRES between 2004 and 2012. The spectral resolutions are in the range $22,000\lesssim R\lesssim 71,000$ for VLT-UVES and $36,000\lesssim R\lesssim 103,000$ for Keck-HIRES. 

We exclude spectra that have low signal-to-noise ratios, roughly SNR$\lesssim2$ per pixel, or do not have spectral coverage at important wavelengths from $\sim$1549 to $\sim$1300 \AA\ that are most likely to have \civ\ outflow lines. We also exclude spectra containing broad BALs or FeLoBALs that substantially cover these same important wavelengths and thus limit our search for mini-BALs. A fraction of the KODIAQ spectra, roughly 5 to 10 percent, needed to be excluded because they contain many spectral pixels without data due to problems in data reductions.

\begin{figure}
\centering
\includegraphics[width=0.5\textwidth]{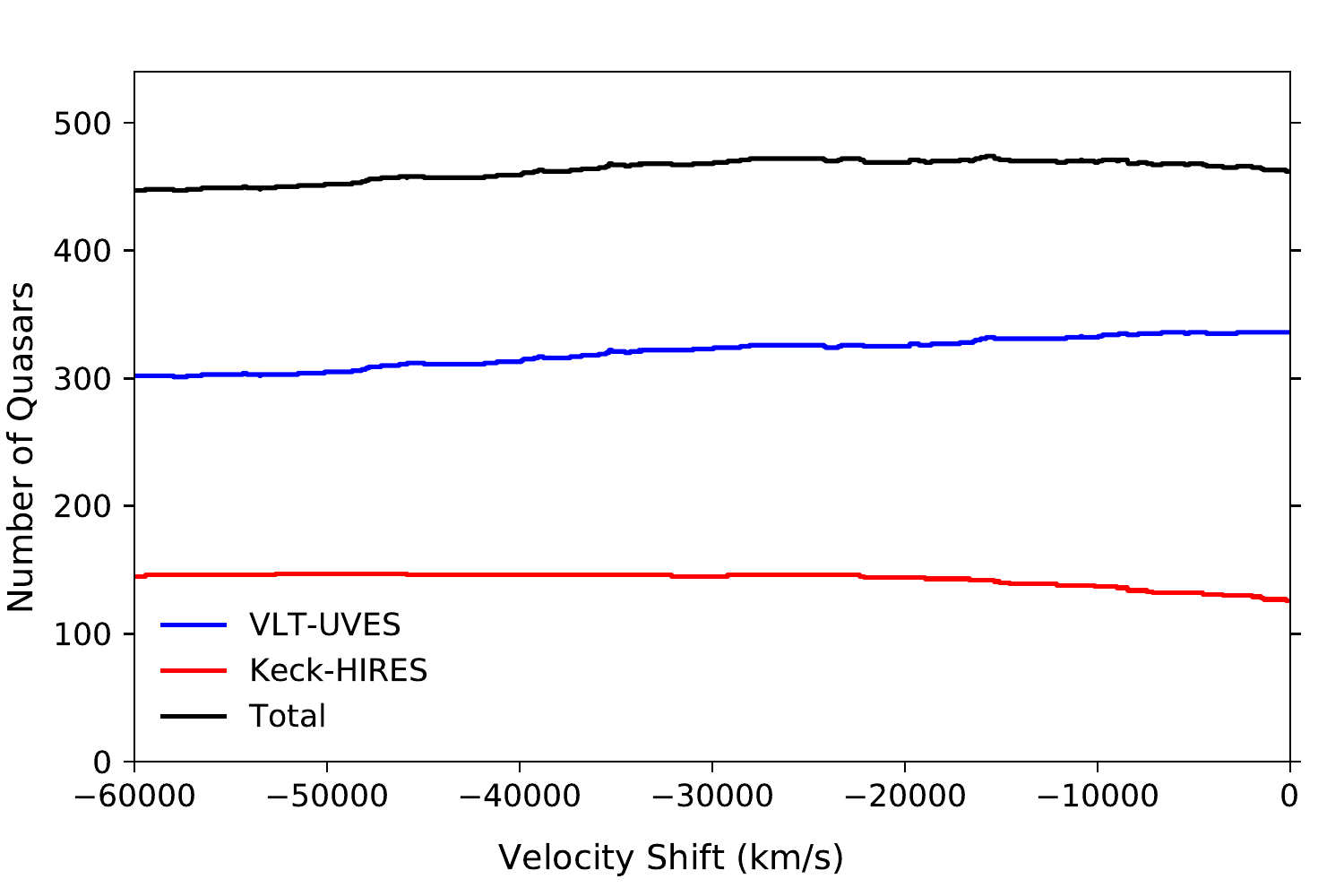}
\caption{Number of quasars measured at different velocity shifts in the VLT-UVES, Keck-HIRES and combined datasets. The velocities are relative to \civ\ \lam 1548 in the quasar rest frame. The negative velocities indicate blueshifts.\label{fig:n_v}}
\end{figure}

Our final samples with suitable spectra include 450 quasars (450 spectra) from UVES and 150 quasars (189 spectra) from HIRES. \Cref{fig:n_v} shows the numbers of quasars with spectral coverage as a function of velocity shift relative to \civ\ \lam 1548 in the quasar frame. The quasar redshifts used here and throughout the remainder of this paper are as provided by \citep{Murphy19} and \citep{OMeara15}. Note that the number of quasars in the UVES sample peaks at only $\sim$330 in \Cref{fig:n_v} because some of the spectra do not cover the full velocity range shown in this plot.

\subsection{The \civ\ mini-BAL catalog}

Our goal is to create a complete catalog of moderately broad \civ\ outflow lines (narrower than BALs) across the velocity range plotted in \Cref{fig:n_v}. Again, for convenience, we refer to all of these lines as mini-BALs. We require, specifically, that the lines have widths in the range $70\lesssim$~FWHM(1548)~$\lesssim 2000$ \kms\ (for just the \lam 1548 component of the \civ\ doublet) at velocity shifts v$\,\lesssim -4000$ \kms\ in the quasar rest frame (where negative values indicate a blueshift). The upper limit on FWHM(1548) excludes BALs based on the common definition of BALs in the ``balnicity" index \citep{Weymann91}. This maximum FWHM threshold is a soft limit, and we specifically search for broader outflow lines at large velocities $\lesssim -30,000$ \kms\ because those systems are not counted in previous large BAL surveys \citep[e.g.,][and refs. therein]{Trump06, Paris17}. The minimum FWHM threshold is needed to avoid large numbers of narrow absorption lines that form in cosmologically intervening gas or galaxies unrelated to the quasars. We chose the specific threshold FWHM $\gtrsim$ 70 \kms\ after careful inspection of the spectral dataset; it is large enough to easily identify broad smooth line troughs characteristic of outflows while also being large enough to reject narrower lines that are more likely to be intervening and unrelated to the quasars. We are also guided by previous \civ\ NAL studies that attempt to identify outflow systems based on partial covering signatures \citep{Misawa07, simonphd}. Those studies show that the lines \textit{without} partial covering, and therefore unclassified but more likely to be intervening, have FWHM that are typically $\lesssim$25 \kms . We impose a minimum velocity shift to avoid narrow ``associated'' absorption lines (AALs), near the quasar redshifts, that can appear in complex blends or have unusual absorption troughs that, in spite of broad widths, have square profiles that do not appear BAL-like and are not likely to form outflows like BALs and mini-BALs \citep[see][for an example of the type broad low-velocity system we avoid with this velocity cut]{Hamann01}. We chose the specific cutoff at $\sim -4000$ \kms\ to avoid specific cases like this evident in our dataset. 

We construct the mini-BAL catalog by first visually inspecting every spectrum in our sample to select mini-BAL candidates based on the criteria above. We then fit the line profiles and search for other lines at the same velocity shift. \civ\ \lam 1548,1551 absorption is usually easy to identify based on the doublet separation of 498 \kms . For broad systems with unresolved doublets, we identify \civ\ absorption by its velocity shift in the quasar frame and the presence or absence of other mini-BALs at the same shift \citep[see][for more discussion]{Paola08, Hamann13}. We find that the continuum fitting performed by \cite{OMeara15} and \cite{Murphy19} to produce normalized spectra can remove real mini-BALs with broad profiles and introduce anomalous broad absorption dips, e.g., in the wings or peaks of broad emission lines where there spectral slopes change dramatically. We avoid these problem for the VLT-UVES sample by visually inspecting both the normalized and unnormalized spectrum of every quasar. For the Keck-HIRES sample, only normalized spectra are available in the KODIAQ data release, but we check all of the broad mini-BAL candidates by inspecting the unnormalized spectra available as quick-look reductions in the online Keck Observatory Archive. This allows us to reject spurious mini-BALs identified as candidates in the normalized spectra. However, we do not inspect the unnormalized Keck-HIRES spectra to select mini-BAL candidates and, therefore, some broad mini-BALs might be missed in that sample if they are removed by the normalization fits. For quasars with BALs in their spectra, we include mini-BALs in our catalog only if they have clearly distinct features, e.g., residing fully outside the BAL troughs or they are clearly distinct features in the weak wing of a BAL trough. These are subjective choices affecting a small number mini-BALs, which we discuss further below. 

\begin{figure}
\centering
\includegraphics[width=0.5\textwidth]{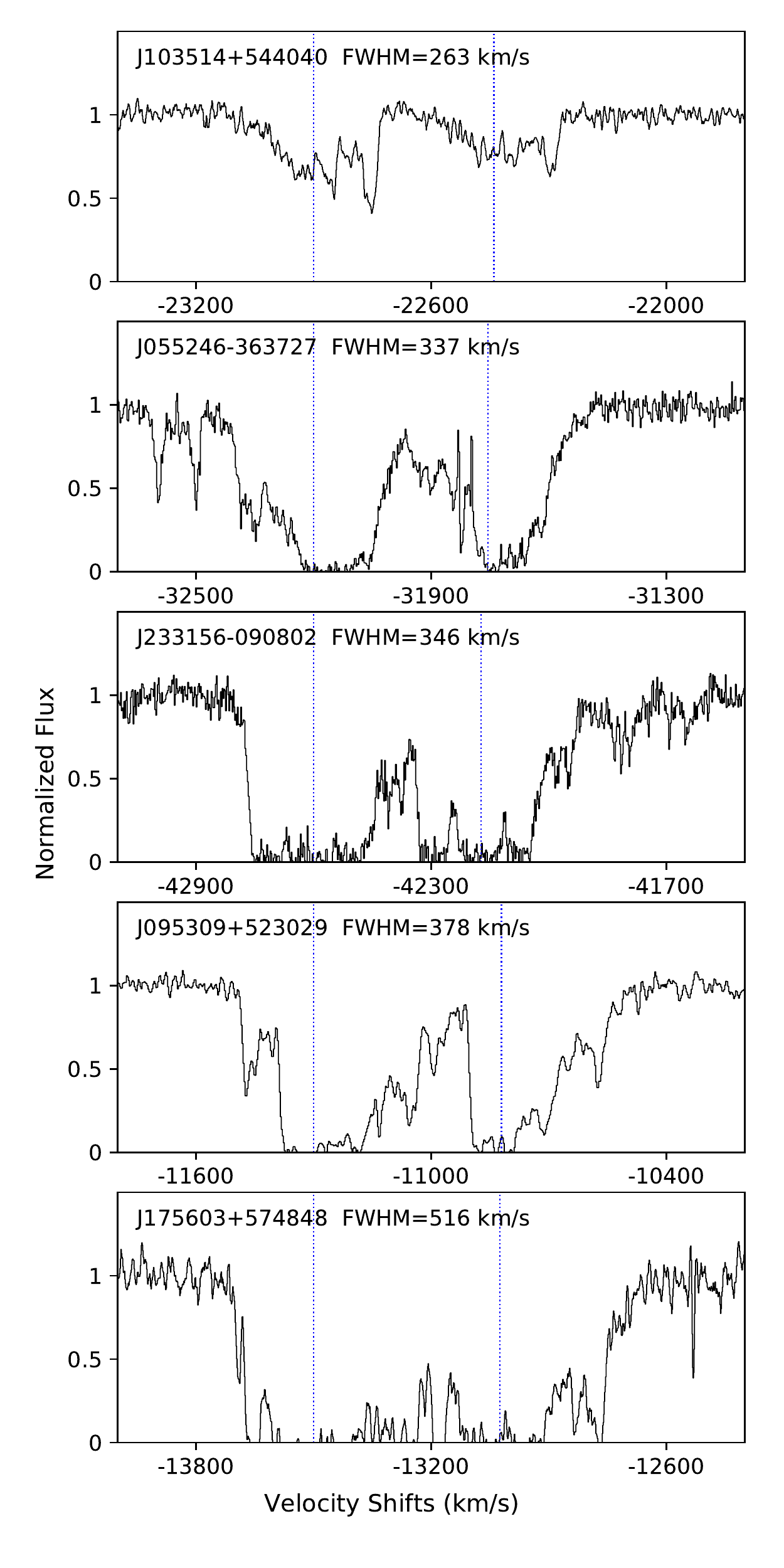}
\vskip -12pt
\caption{Normalized spectra showing examples of \civ\ absorption lines in DLAs or LLSs, plotted on a velocity scale relative to the quasar redshift \citep{OMeara15, Murphy19}. The blue dotted lines mark the approximate centroids of the two \civ\ doublet components. The lines plotted all have FWHM$>250$ \kms , illustrating the overlap with mini-BALs in FWHM parameter space, but their membership to DLA or Lyman Limit systems is confirmed by measurements of \lya\ and other lines at the same redshift. \label{fig:dla}}
\end{figure}

Next we inspect every candidate mini-BAL system for indications that it might be an unrelated intervening absorption line system. The most obvious contaminants we find among our mini-BAL candidates are \civ\ lines that belong to damped \lya\ (DLA) or Lyman limit systems (LLSs), which form in intervening galaxies or extended halos and can have \civ\ lines broad enough to overlap with our mini-BAL selection \citep[e.g.,][]{Prochaska15, Berg15, Prochaska08}. \Cref{fig:dla} shows examples of \civ\ doublets in DLA systems that have FWHM(1548)~$>250$ \kms . We automatically reject all \civ\ systems if they have broad and deep \lya\ lines indicative of DLAs or LLSs within our spectral coverage. If the Lyman lines are not covered, we search for low-ionization metal lines that are common in DLAs and LLSs, such as \siii\ \lam 1260,1527, \oi\ \lam 1304, \cii\ \lam 1335, and \feii\ \lam 1608. We also examine the \civ\ line profiles and reject systems that resemble the \civ\ lines found in DLA systems, e.g., with deep and narrow profiles with steep vertical sides, or highly asymmetric and clearly composed of blended narrow components. Overall, we reject systems with any reasonable possibility of being intervening so that our final mini-BAL catalog is free of contamination. This conservative approach means that some real outflow mini-BALs (with unusual profiles or other characteristics) could be excluded from our survey. However, the number of mini-BAL candidates rejected because they might belong to a DLA system, e.g., not obviously in DLAs by the criteria above, is small compared to the total number of mini-BALs in our final catalog. Thus, we estimate that our catalog is $>90$\% complete for the ensemble selection criteria described above and for line strengths greater than some nominal sensitivity thresholds. The quantitative line data provided by our fitting analysis (Section 3) indicates that the sensitivity limit of our survey is conservatively REW(1548)~$\gtrsim 0.03$ \AA\ for lines with FWHM(1548)~$\sim 100$ \kms\ (or roughly double that REW for lines that are twice as broad, etc.). 

\section{Line Profile Fits}

The first step in our fitting procedure is to carefully examine the continuum placement provided with the UVES and HIRES archive spectra in the vicinity of every line we will fit. In some cases (e.g., for broad mini-BALs), the continuum extrapolated across the absorption lines appears poorly constrained or too sharply varying compared to the unnormalized spectrum. In those cases, we redo the continuum fit using a simple power law constrained locally near the line. We fit every candidate \civ\ mini-BAL using a Gaussian optical depth profile given by,
\begin{equation}
\label{eq:1}
\tau_{v}=\tau_{0}e^{-{(v-v_0)}^2/b^2},
\end{equation}
where $\tau_v$ is the optical depth at velocity $v$, $\tau_{0}$ is the line center optical depth, $v_0$ is the line center velocity, and $b$ is the Doppler parameter. The doublet components in \civ\ \lam 1549,1551 are fit simultaneously at the same velocity shift and $b$ value. We account for partial covering in our line fits, but we assume for simplicity that the background light source has a spatially uniform brightness and the absorbing medium is homogeneous, with the same optical depth along every sightline, so the observed intensity at velocity $v$ is given by 
\begin{equation}
\label{eq:2}
\frac{I_{v}}{I_{0}}=1-C_v+C_ve^{-\tau_{v}},
\end{equation}
where $I_{0}$ is the continuum intensity, $I_{v}$ is the measured intensity at velocity $v$, and $C_v$ is the covering fraction of the absorbing medium across the emission source such that $0 < C_v \leq1$ \citep{Ganguly99, Hamann97b, Barlow97b}. We also assume that the covering fraction is constant with velocity across the mini-BAL profiles, i.e., $C_v=C_0$. This is justified by our fits below showing that, in general, the effects of any velocity dependence in $C_v$ are negligible compared to the overall value of the covering fraction captured by the constant $C_0$. 

An important caveat to \Cref{eq:2} is that quasar outflows are often spatially inhomogeneous, with a range of column densities and line optical depths across the projected area of the emission source \citep[see Section 5, also][and refs. therein]{Barlow97b, deKool02, Hamann01, Hamann04, Arav02, Arav05, Hamann19a, Hamann19b}. This situation leads to optical depth-dependent covering fractions, as derived from \Cref{eq:2}. For example, if the absorber spans a range of optical depths from thick to thin at different spatial locations, then 1) stronger transitions will yield larger covering fractions because they are optically thick over larger areas, and 2) the relative strengths of different absorption lines (as in the \civ\ doublet) will depend on both the magnitude and specific spatial distribution of the line optical depths across the emission source. The actual spatial distributions cannot be determined from a simple doublet analysis. However, previous studies have shown that, even with inhomogeneous absorbers, \Cref{eq:2} yields useful estimates of the $\tau_0 \gtrsim 1$ covering fractions in different lines as well as a spatially-averaged estimate of the optical depths \citep{Hamann04, Arav05}.  

\Cref{fig:mbal_all} shows our fits to all of the \civ\ mini-BALs included in our final catalog (described in Section 4 below). We measure $v$, $\tau_0$ and $C_0$ from \Cref{eq:1,eq:2} by fitting the \civ\ doublet lines simultaneously, with their $\tau_0$ values fixed to the $\sim$2:1 ratio of their transition strengths. Most of the fits are straightforward. However, the data can present difficulties in three general ways: 1) the \civ\ doublet lines in the mini-BAL are broad enough to be fully blended together, 2) the mini-BAL is badly blended with another mini-BAL or unrelated lines, and 3) the mini-BALs have complex profiles not well-characterized by a Gaussian optical depth function. We deal with each of these complicated cases as follows:

\textit{Case 1)} We fit broad blended mini-BALs doublets assuming the lines have a $\sim$1:1 optical depth ratio. This is not physical but we find that it yields the best fits to the observed profiles and, therefore, the best measurements of the line parameters. If the broad blended mini-BALs have shallow or moderate depths (not reaching zero intensity) and rounded, roughly symmetric profiles, our fits yield firm lower limits on both $\tau_0$ and $C_0$ (equal to the line depth below the continuum). If the broad blended mini-BALs, instead, have roughly flat-bottom troughs, good fits require $\tau_0\gg 1$ to saturate the line cores. Thus large optical depths with partial covering are indicated even though the doublet ratio is not available. In these latter situations, we adopt the measured line depths as the covering fraction reported in our catalog because, in the optically thick limit, \Cref{eq:2} simplifies to $C_0\approx1-I_{v}/I_{0}$. 

\textit{Case 2)} We fit all \civ\ mini-BALs simultaneously if there are two or more such systems blended together. Examples of this are the low-velocity mini-BALs in J131215+423900 and J211654-433234 (\Cref{fig:mbal_all}). We identify and fit distinct mini-BALs in these blends only if they are clearly distinct by visual inspection, e.g., with well-measured absorption minima separated by $\gtrsim100$ \kms . In some cases, we fit one \civ\ doublet to what is probably a more complex blend of mini-BALs. This procedure is somewhat subjective, but we are deliberately conservative to avoid over-counting the number mini-BALs in our dataset. For \civ\ mini-BALs blended with unrelated absorption lines (at some other redshift), we simply mask out the spectral regions containing the unrelated lines before fitting. This is generally straightforward. However, there are rare cases where the fits are poorly constrained because the masked regions are large due to BALs near the mini-BALs (e.g., for the mini-BALs at $-$4549 \kms\ in J235702-004824 and roughly $-$7000 \kms\ in J221531-174408 in \Cref{fig:mbal_all}) or because part of the spectrum is missing due to gaps in wavelength coverage between the echelle orders (e.g., for the mini-BAL at $-$4544 \kms\ in J102325+514251 in \Cref{fig:mbal_all}). 

\textit{Case 3)} We fit mini-BALs with complex/non-Gaussian profiles using the minimum number of Gaussian components to provide a good fit. Again, our goal is to avoid over-counting the mini-BALs in our study. As noted for Case 2 above, distinct components must have well-measured absorption minima separated by $\gtrsim100$ \kms . The fits to these complex profiles are not ideal, but they achieve our goals of i) conservatively counting the number of mini-BALs in the spectra and ii) measuring the basic line properties. These complex profile cases range from moderately non-Gaussian, such as the mini-BAL at $-11,966$ \kms\ in J024221+004912, the one at $-$10,018 \kms\ in J104642+053107, the one at $-$13,061 \kms\ in J211654-433234, to highly-structured blend of mini-BALs at $-$21,767 \kms\ of J104032-272749 (\Cref{fig:mbal_all}). 

All of the broad and blended mini-BALs are flagged in the catalog notes in \Cref{tab:mBAL_uves,tab:mbal_keck}. Additional notes on specific cases with complex blends, under-counted outflow lines, or problematic/uncertain continuum placements are provided in Appendix A. 

\section{Results \& Analysis}

\Cref{tab:mBAL_uves,tab:mbal_keck} provide our final catalog of \civ\ mini-BALs in the VLT-UVES and Keck-HIRES datasets, respectively. The measured quantities in these tables are the velocity shift, $v$, covering fraction $C_0$, rest equivalent widths, REW, given separately for the \civ\ \lam 1548, 1551 doublet lines, and the FWHM for \civ\ \lam 1548 only (it is the same for both lines). The uncertainties listed for most of the parameters are 1$\sigma$ errors output by the fitting software; these are due mainly to pixel-to-pixel noise fluctuations in the spectra. The tables also include Notes to indicate specific cases with potentially larger uncertainties due blends, uncertain continuum fits, or other problems. These notes are explained in the table caption. One of those designations is ``cmplx" to indicate that the recorded mini-BALs belong to a complex of related outflow lines. We define an absorption-line complex as three or more \civ\ mini-BALs that appear to be related based on similar profiles and roughly similar velocity shifts ($\lesssim3000$ \kms\ apart). These complexes sometimes include outflow lines not in our mini-BAL catalog because they are too narrow or at velocities v$\,> -4000$ \kms\ (see Section 4.3 for specific examples and discussion). 

\begin{figure*}
\centering
\includegraphics[width=1\textwidth]{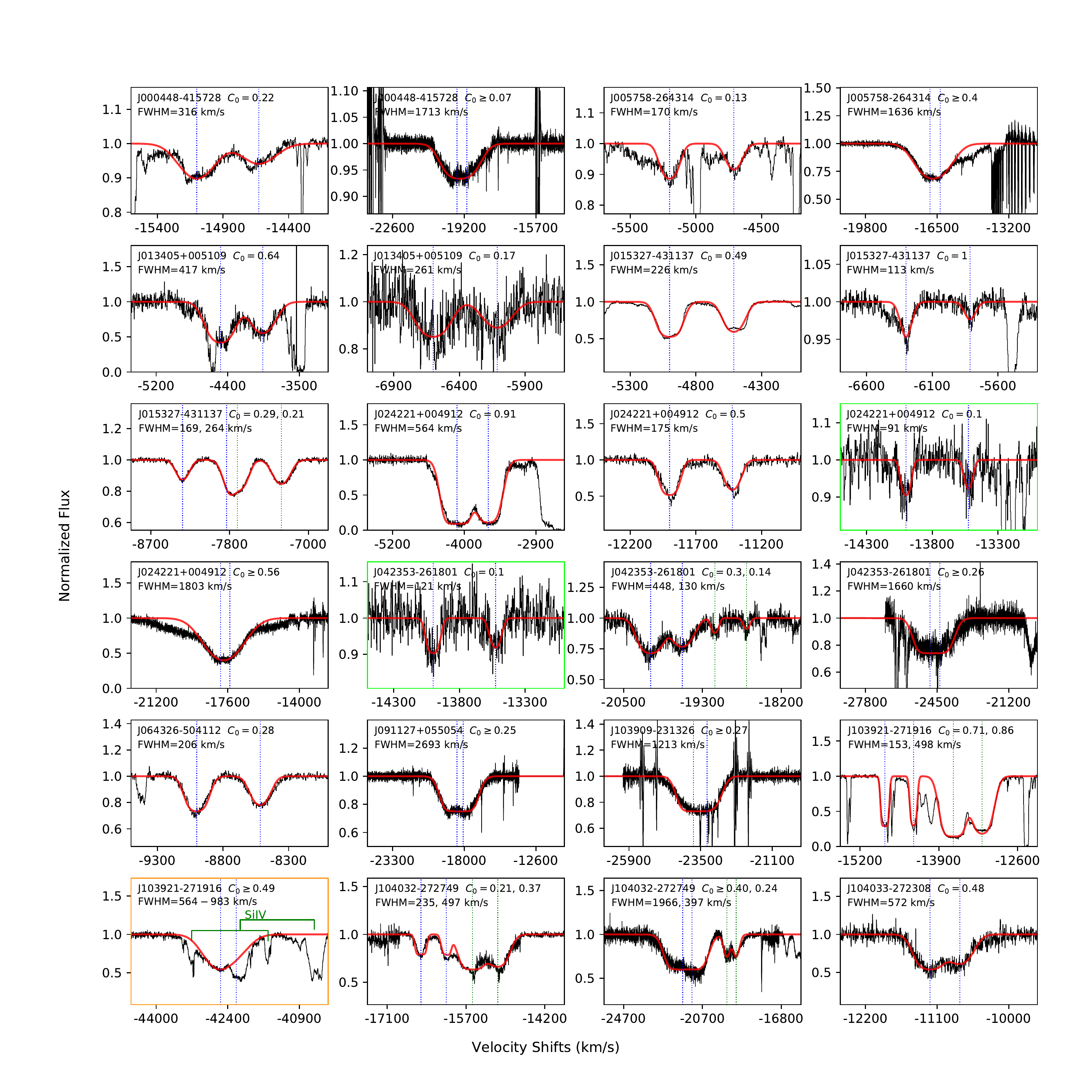}
\caption{Normalized \civ\ mini-BAL profiles plotted on a velocity scale relative to the quasar emission-line redshift. The spectra are shown in black. Our fits to the lines are shown in red. The dash lines with different colors indicate separate mini-BAL systems, and we fit the separate mini-BAL systems in each panel simultaneously. The green boxes highlight the systems with definite $C_0<0.1$, where we exclude the ones with only lower limits and the ones with uncertain measurements, and the orange boxes highlight the systems with extremely high speeds, $v<-30,000$ km/s. Observing dates are given (in blue text) if quasars observed more than once. The velocities pertain to the short-wavelength lines in the doublets.\label{fig:mbal_all}}
\end{figure*}

\begin{figure*}
\centering
\setcounter{figure}{2}
\includegraphics[width=1\textwidth]{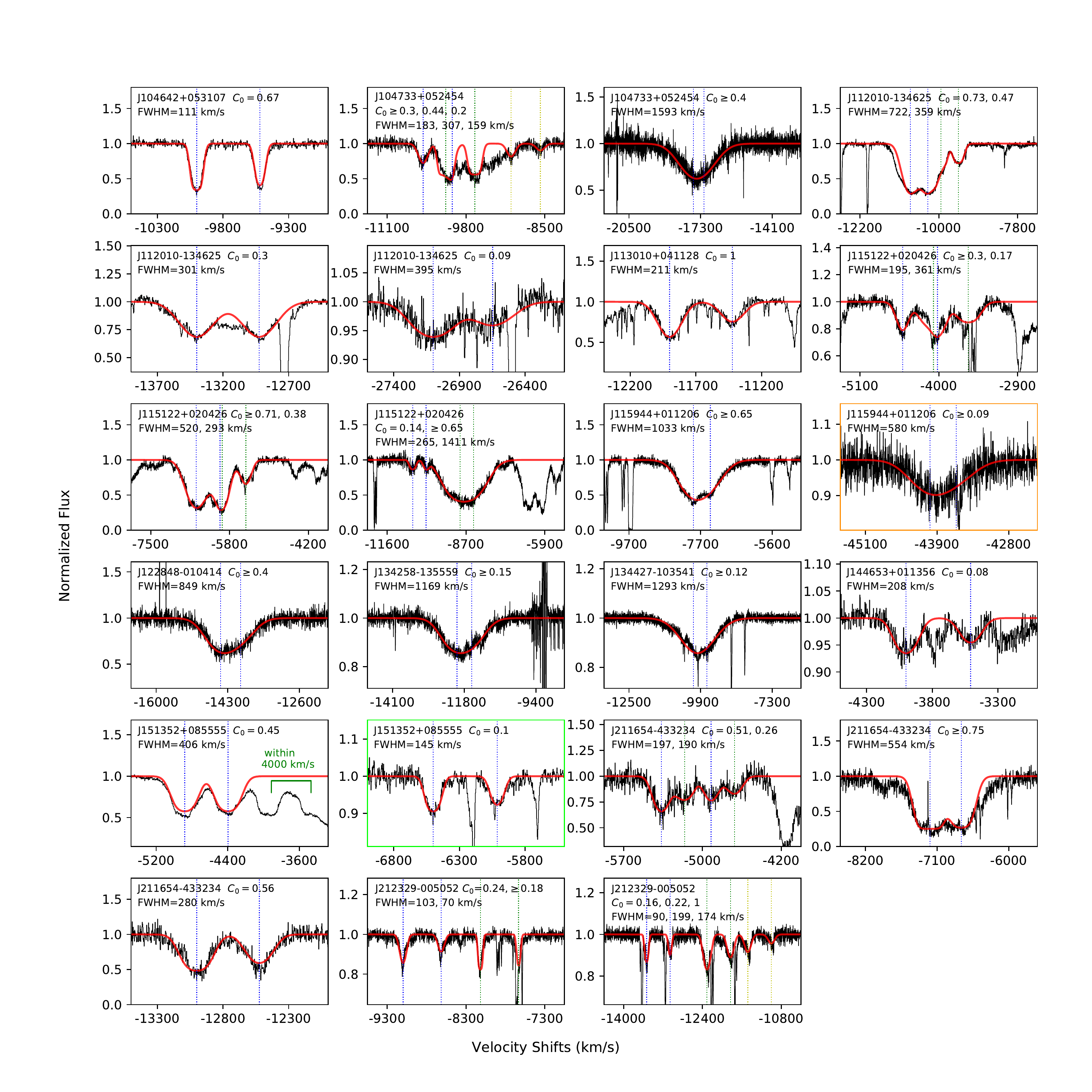}
\caption{\textit{continued.}}
\end{figure*}

\begin{figure*}
\centering
\setcounter{figure}{2}
\includegraphics[width=1\textwidth]{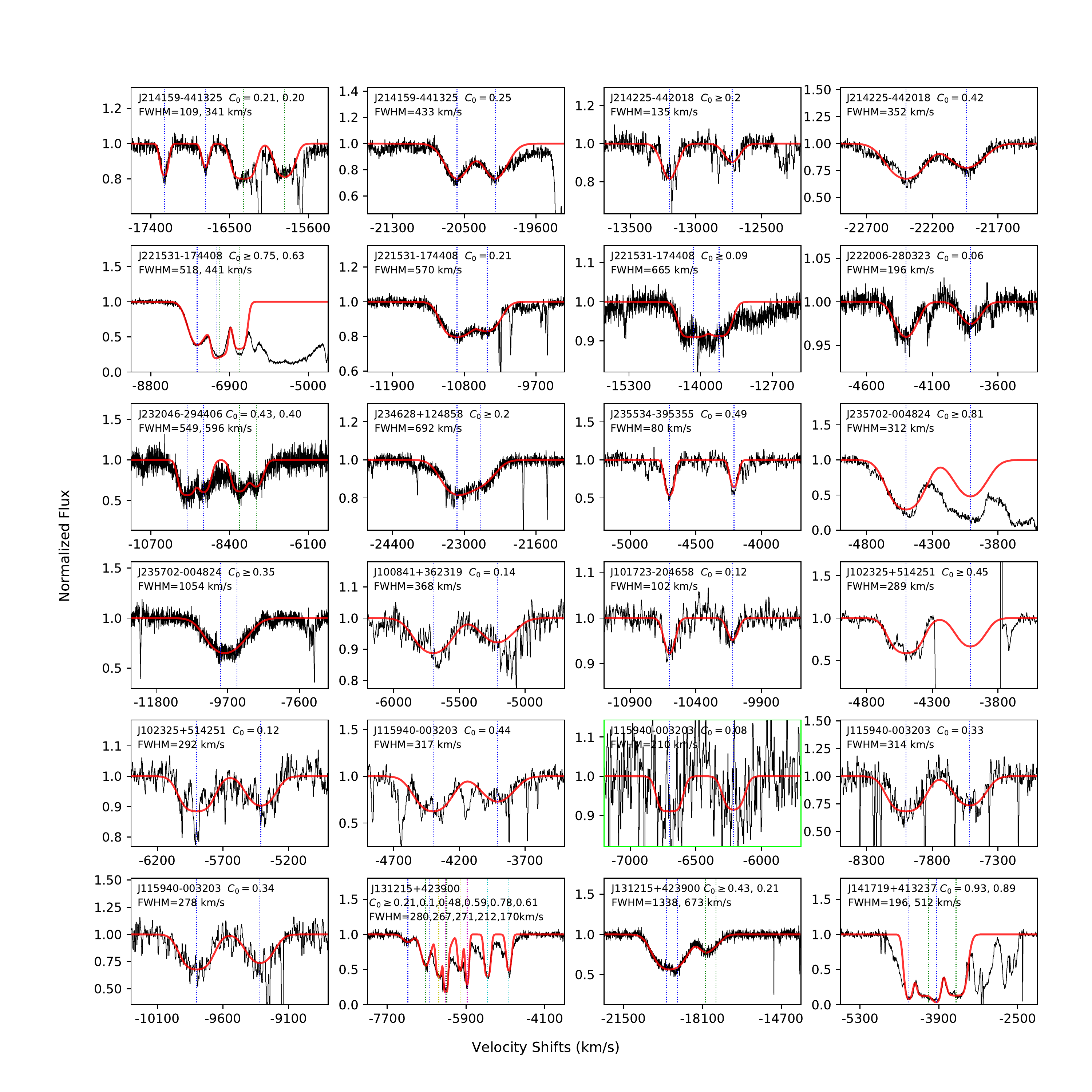}
\caption{\textit{continued.}}
\end{figure*}

\begin{figure*}
\centering
\setcounter{figure}{2}
\includegraphics[width=1\textwidth]{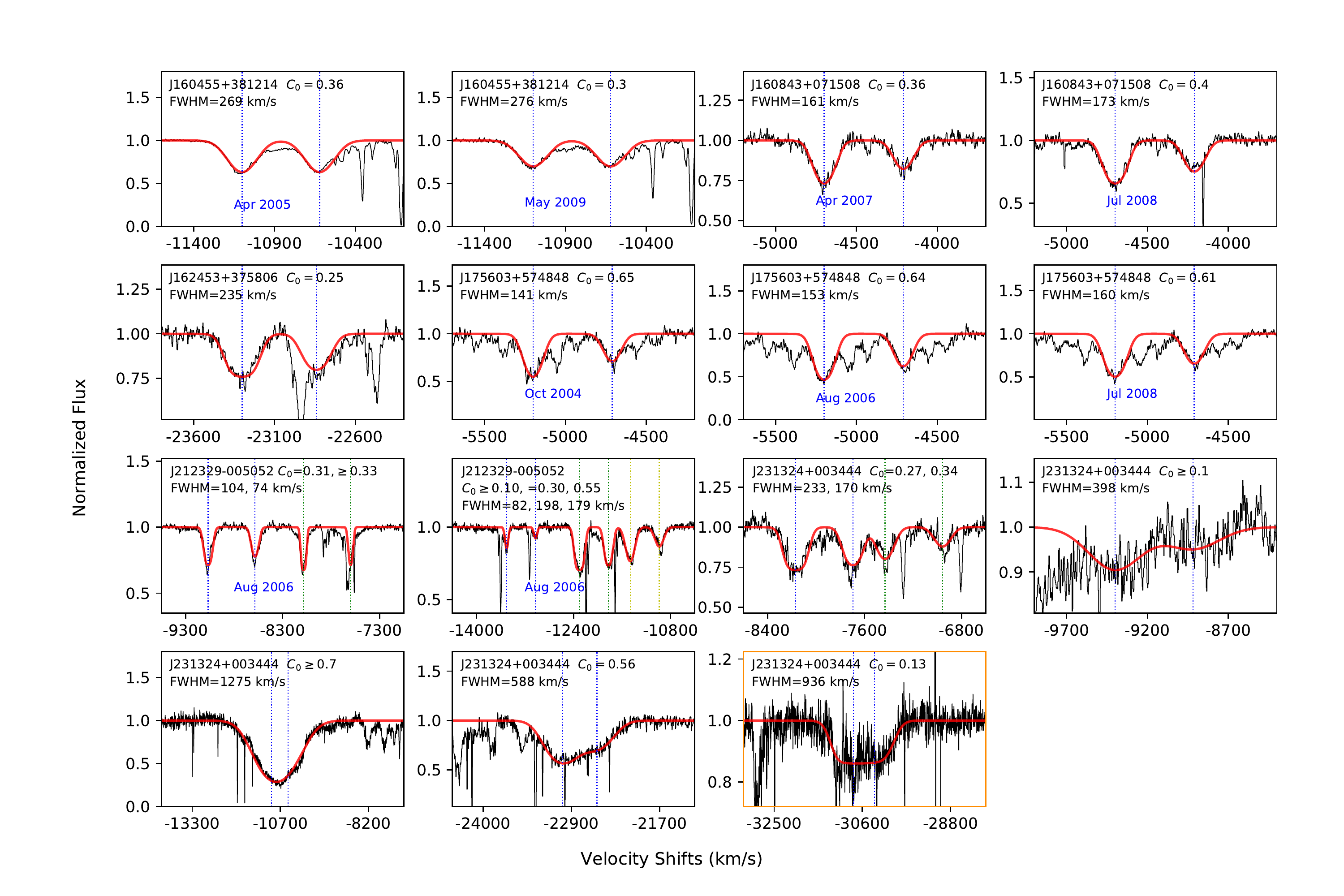}
\caption{\textit{continued.}}
\end{figure*}

\clearpage

\startlongtable
\begin{deluxetable*}{cccccccc}
\tablecaption{\civ\ mini-BALs in VLT-UVES. Columns show the quasar names, emission-line redshifts, $z_{em}$, velocity shifts, $v$, of the absorption-line centroid from $z_{em}$, rest equivalent widths, REWs, for \lam 1548 and \lam1551 in the \civ\ doublet, FWHMs for the separate \civ\ lines, covering fractions, $C_0$, inferred from the line fits, and Notes on individual features: `broad' = broad profiles where the \civ\ doublet components are blended together and indistinguishable as separate troughs; `bl1' =  asymmetric non-gaussian profile that we fit approximately with a single gaussian component; `bl2' = blended with distinct unrelated lines (if the blended system is also a measured mini-BAL, we identify the blended system by its velocity shift in the notation bl2-xxxx); `bl-BAL' = in the wing of a much broader \civ\ BAL; `cmplx' = appears to belong to a complex of related outflow lines, see Section 4.0; `BALQSO' = the quasar has a \civ\ BAL; `unc' = the fit data listed have uncertainties larger than indicated due to blends or uncertain continuum placement; `mBAL?' = the fitted feature appears to be a mini-BAL but it has some characteristics resembling \civ\ lines in DLA systems; `var' = variability is identified between two or more VLT-UVES and/or Keck-HIRES observations included in our study via visual inspection (these variable systems have multiple table entries distinguished in the Notes by the observation dates).\label{tab:mBAL_uves}}
\tabletypesize{\scriptsize}
\tablehead{
\colhead{QSO} & \colhead{$z_{em}$} & \colhead{$v$} & \colhead{REW(\lam1548)} & \colhead{REW(\lam1551)} & \colhead{FWHM(\lam1548)} & \colhead{$C_0$} & \colhead{Notes} \\ 
\colhead{} & \colhead{} & \colhead{(\kms)} & \colhead{(\AA)} & \colhead{(\AA)} & \colhead{(\kms)} &\colhead{} & \colhead{}
}  
\startdata
J000448-415728 & 2.760 & -15136 & $0.17\pm0.03$ & $0.10\pm0.02$ & $316.4\pm41.0$ & $0.22\pm0.04$ & bl1\\ 
  &   & -19565 & $0.52\pm0.01$ & $0.35\pm0.02$ & $1712.5\pm67.0$ & $\gtrsim0.06$ & broad \\
J005758-264314 & 3.655 & -5273 & $0.10\pm0.01$ & $0.07\pm0.01$ & $169.5\pm19.4$ & $0.13\pm0.01$ & bl2 \\ 
  &   & -16876  & $2.14\pm0.07$ & $1.35\pm0.06$ & $1636.1\pm56.3$ & $\gtrsim0.34$ & broad, bl2, unc \\
J013405+005109 & 1.520 & -4501 & $1.27\pm0.04$ & $0.86\pm0.05$ & $416.9\pm19.5$ & $0.64\pm0.02$ & bl2 \\
 & & -6666 & $0.22\pm0.03$ & $0.15\pm0.03$ & $260.8\pm48.3$ & $0.17\pm0.02$ & unc \\
J015327-431137 & 2.740 & -5080 & $0.55\pm0.01$ & $0.40\pm0.04$ & $226.0\pm6.0$ & $0.49\pm0.01$ & cmplx\\ 
 &  & -6402 & $0.03\pm0.01$ & $0.02\pm0.00$ & $113.1\pm15.0$ & $1.0$ &  bl2, unc, cmplx\\
 &  & -7789 & $0.27\pm0.00$ & $0.18\pm0.00$ & $264.1\pm5.4$ & $0.21\pm0.00$ & bl2-8387, cmplx\\ 
 &  & -8387 & $0.12\pm0.02$ & $0.09\pm0.02$ & $169.3\pm26.0$ & $0.29\pm0.06$ & bl2-7789, cmplx\\ 
J024221+004912 & 2.068  & -4189 & $2.68\pm0.02$ & $2.20\pm0.07$ & $564.3\pm10.7$ & $0.91\pm0.00$ & \\ 
  &   & -11966 & $0.43\pm0.01$ & $0.32\pm0.09$ & $174.7\pm8.7$ & $0.50\pm0.01$ & \\
  &   & -14063 & $0.05\pm0.01$ & $0.03\pm0.02$ & $90.7\pm21.3$ & $0.10\pm0.02$ & \\ 
  &   & -18003 & $5.09\pm0.48$ & $2.99\pm0.32$ & $1802.5\pm61.8$ & $\gtrsim0.56$ & broad, bl2, unc  \\
J042353-261801 & 2.277 & -14013 & $0.06\pm0.01$ & $0.04\pm0.02$ & $121.1\pm25.2$ & $0.10\pm0.01$ & \\ 
  & & -19203 & $0.08\pm0.02$ & $0.05\pm0.02$ & $129.6\pm34.2$ & $0.14\pm0.03$ & \\
  &   & -20145 & $0.65\pm0.03$ & $0.46\pm0.03$ & $448.2\pm26.8$ & $0.30\pm0.01$ & \\  
  &   & -24869 & $2.24\pm0.23$ & $1.81\pm0.30$ & $1659.9\pm407.6$ & $\gtrsim0.26$ & broad, unc \\
J064326-504112 & 3.090 & -9017 & $0.29\pm0.01$ & $0.20\pm0.03$ & $206.1\pm7.0$ & $0.28\pm0.01$ & \\ 
J091127+055054 & 2.793 & -18544 & $3.02\pm0.03$ & $2.73\pm0.07$ & $2692.6\pm70.1$ & $\gtrsim0.25$ & broad \\
J103909-231326 & 3.130 & -23808 & $1.69\pm0.05$ & $1.36\pm0.11$ & $1213.0\pm68.2$ & $\gtrsim0.27$ & broad\\
J103921-271916 & 2.230 & -13684 & $2.22\pm0.02$ & $1.79\pm0.19$ & $497.6\pm7.6$ & $0.86\pm0.00$ & cmplx, bl2\\
  &   & -14816 & $0.57\pm0.01$ & $0.48\pm0.18$ & $153.3\pm7.9$ & $0.71\pm0.01$ & cmplx, bl2\\
 & & -42626 & \multicolumn{2}{c}{$0.94-2.42$} & $564.2-982.7$ & $\gtrsim0.49$ & broad, bl2 with \siiv, unc \\
J104032-272749 & 2.320 & -15623 & $0.98\pm0.02$ & $0.78\pm0.03$ & $496.9\pm23.5$ & $0.37\pm0.01$ & bl2-16568, cmplx\\ 
  &    & -16568 & $0.26\pm0.01$ & $0.22\pm0.01$ & $235.0\pm18.8$ & $0.21\pm0.01$ & bl2-15623, cmplx\\ 
  &   & -19506 & $0.62\pm0.02$ & $0.53\pm0.08$ & $396.5\pm29.4$ & $0.24\pm0.01$ & cmplx\\    
  &   & -21767 & $4.04\pm0.06$ & $3.36\pm0.15$ & $1966.0\pm63.7$ & $\gtrsim0.40$ & broad, cmplx, bl1, unc \\
J104033-272308 & 1.937 & -11259 & $1.37\pm0.04$ & $0.98\pm0.08$ & $571.6\pm31.3$ & $0.48\pm0.01$ & bl-BAL, BALQSO\\ 
J104642+053107 & 2.698 & -10018 & $0.38\pm0.01$ & $0.29\pm0.10$ & $110.5\pm3.6$ & $0.67\pm0.01$ & \\ 
J104733+052454 & 1.334 & -9100 & $0.16\pm0.03$ & $0.08\pm0.02$ & $158.9\pm11.3$ & $\gtrsim0.2$ & bl2, cmplx, unc\\ 
 & & -10172 & $0.73\pm0.02$ & $0.63\pm0.03$ & $307.0\pm23.3$ & $0.44\pm0.01$ & bl2-10552, cmplx\\ 
  &  & -10552 & $0.27\pm0.04$ & --- & $183.1\pm12.7$ & $\gtrsim0.3$ &  bl2-10172, cmplx\\ 
  & & -17662  & $2.40\pm1.04$ & $1.27\pm0.60$ & $1593.1\pm448.4$ & $\gtrsim0.4$ & broad\\
J112010-134625 & 3.958 & -10002 & $0.77\pm0.02$ & $0.50\pm0.05$ & $359.2\pm12.1$ & $0.47\pm0.01$ & bl2-10806, BALQSO\\
  &   & -10806 & $2.74\pm0.02$ & $2.22\pm0.13$ & $721.7\pm16.0$ & $0.73\pm0.00$ & bl2-10002, BALQSO \\
  &   & -13416 & $0.48\pm0.01$ & $0.39\pm0.07$ & $300.6\pm19.6$ & $0.30\pm0.00$ & unc, bl-BAL, BALQSO\\ 
  & & -27118 & $0.13\pm0.02$ & $0.08\pm0.02$ & $395.4\pm67.5$ & $0.09\pm0.02$ & unc, bl-BAL, BALQSO\\
J113010+041128 & 3.930 & -11975 & $0.49\pm0.14$ & $0.27\pm0.09$ & $211.3\pm43.5$ & $1.0$ & unc, BALQSO\\ 
J115122+020426 & 2.401 & -4107 & $0.32\pm0.02$ & $0.24\pm0.02$ & $360.9\pm41.6$ & $0.17\pm0.01$ & bl2-4529, bl2, cmplx\\
& & -4529 & $0.24\pm0.17$ & $0.13\pm0.10$ & $194.7\pm122.4$ & $\gtrsim0.30$ & bl2-4107, cmplx\\
 &  & -6046 & $0.75\pm0.07$ & $0.47\pm0.06$ & $292.6\pm28.7$ & $\gtrsim0.38$ & bl1, bl2-6578, cmplx\\ 
  &   & -6578 & $1.88\pm0.19$ & $1.11\pm0.14$ & $520.2\pm48.3$ & $\gtrsim0.71$ & bl1, bl2-6046, cmplx\\ 
  &   & -9024 & $4.01\pm0.10$ & $2.60\pm0.10$ & $1411.6\pm45.7$ & $\gtrsim0.65$ & broad, cmplx, unc \\
  &   & -10703 & $0.18\pm0.02$ & $0.13\pm0.03$ & $265.2\pm36.1$ & $0.14\pm0.01$ & cmplx, unc\\
J115944+011206 & 2.000 & -7917 & $2.72\pm0.23$ & $1.55\pm0.15$ & $1033.2\pm82.4$ & $\gtrsim0.65$ & broad\\
  & & -44093 & $0.28\pm0.02$ & $0.22\pm0.07$ & $579.6\pm69.0$ & $\gtrsim0.09$ & broad, unc\\
J122848-010414 & 2.655 & -14474 & $1.54\pm0.18$ & $0.88\pm0.12$ & $849.1\pm91.5$ & $\gtrsim0.40$ & broad\\
J134258-135559 & 3.190 & -12070 & $0.78\pm0.07$ & $0.48\pm0.05$ & $1168.7\pm108.6$ & $\gtrsim0.15$ & broad\\
J134427-103541 & 2.134 & -10226 & $0.75\pm0.07$ & $0.41\pm0.06$ & $1293.3\pm143.1$ & $\gtrsim0.12$& broad\\
J144653+011356 & 2.206 & -4004 & $0.07\pm0.01$ & $0.04\pm0.01$ & $208.1\pm32.8$ & $0.08\pm0.01$ & cmplx, unc\\ 
J151352+085555 & 2.904 & -4922 & $0.95\pm0.01$ & $0.75\pm0.04$ & $405.7\pm10.4$ & $0.45\pm0.01$ & bl2, cmplx\\ 
  &   & -6564 & $0.07\pm0.01$ & $0.05\pm0.01$ & $144.7\pm28.4$ & $0.10\pm0.01$ & cmplx \\ 
J211654-433234 & 2.053 & -5180 & $0.22\pm0.03$ & $0.14\pm0.03$ & $189.7\pm43.1$ & $0.26\pm0.04$ & bl2-5409, cmplx\\
& & -5409 & $0.36\pm0.07$ & $0.21\pm0.05$ & $197.0\pm40.3$ & $0.51\pm0.10$ & bl2-5180, cmplx\\
  &   & -7274 & $2.16\pm0.05$ & $1.79\pm0.10$ & $554.1\pm37.1$ & $\gtrsim0.75$ & broad, bl1, cmplx\\
  &   & -13061 & $0.76\pm0.03$ & $0.52\pm0.10$ & $280.0\pm16.5$ & $0.56\pm0.02$ & \\ 
J212329-005052 & 2.262  & -8167 & $0.06\pm0.01$ & $0.05\pm0.02$ & $69.7\pm12.2$ & $\gtrsim0.18$ & bl2, cmplx, var (Aug 2008)\\
& & -9148 & $0.07\pm0.02$ & $0.05\pm0.02$ & $102.6\pm17.6$ & $0.24\pm0.05$ & cmplx, var (Aug 2008) \\
  &   & -11521 & $0.06\pm0.02$ & $0.03\pm0.02$ & $174.0\pm52$ & $1.0$ & cmplx, bl1, unc, var (Aug 2008)\\
  &   & -12336 & $0.18\pm0.01$ & $0.11\pm0.03$ & $199.3\pm18.9$ & $0.22\pm0.02$ & cmplx, bl2, var (Aug 2008)\\
  &   & -13568 & $0.06\pm0.01$ & $0.04\pm0.02$ & $89.5\pm11.3$ & $0.16\pm0.02$ & cmplx, bl2, var (Aug 2008)\\
J214159-441325 & 3.170 & -16414 & $0.34\pm0.01$ & $0.25\pm0.05$ & $340.8\pm15.2$ & $0.20\pm0.01$ & bl2, cmplx \\
  &   & -17319 & $0.10\pm0.01$ & $0.07\pm0.04$ & $109.4\pm11.1$ & $0.21\pm0.02$ & cmplx\\ 
  &   & -20573 & $0.50\pm0.01$ & $0.44\pm0.08$ & $432.5\pm19.5$ & $0.25\pm0.01$ & cmplx\\
J214225-442018 & 3.230 & -13239 & $0.14\pm0.10$ & $0.08\pm0.06$ & $135.3\pm43.4$ & $\gtrsim0.20$ & mBAL?, bl1, unc\\
  &   & -22453 & $0.60\pm0.04$ & $0.37\pm0.09$ & $351.6\pm24.0$ & $0.42\pm0.03$ & \\ 
J221531-174408 & 2.217 & -7155 & $1.55\pm0.02$ & $1.34\pm0.02$ & $440.6\pm11.3$ & $\gtrsim0.63$ & bl2-7711, bl-BAL, cmplx, BALQSO\\ 
&  & -7711 & $1.68\pm0.21$ & $0.97\pm0.15$ & $517.8\pm55.6$ & $\gtrsim0.75$ & bl2-7155, bl-BAL, cmplx, BALQSO\\ 
  &   & -10971 & $0.59\pm0.01$ & $0.42\pm0.03$ & $570.0\pm17.6$ & $0.21\pm0.00$ & BALQSO\\
  &   & -14179 & $0.33\pm0.01$ & $0.26\pm0.02$ & $665.0\pm48.9$ & $\gtrsim0.09$ & broad, bl2, unc, BALQSO\\ 
J222006-280323 & 2.406 & -4355 & $0.04\pm0.01$ & $0.03\pm0.01$ & $196.0\pm26.0$ & $0.06\pm0.01$ & unc\\
J232046-294406 & 2.401 & -8148 & $1.17\pm0.05$ & $0.84\pm0.06$ & $595.5\pm43.7$ & $0.40\pm0.01$ & \\
  &   & -9659 & $1.23\pm0.04$ & $0.97\pm0.05$ & $549.3\pm36.4$ & $0.43\pm0.01$ & bl1\\
J234628+124858 & 2.515 & -23164 & $0.65\pm0.05$ & $0.40\pm0.06$ & $691.5\pm58.2$ & $\gtrsim0.20$ & broad\\
J235534-395355 & 1.580 & -4756 & $0.19\pm0.01$ & $0.14\pm0.03$ & $79.7\pm7.7$ & $0.49\pm0.03$ & \\
J235702-004824 & 3.0178 & -4549 & $1.30\pm0.07$ & --- & $312.2\pm32.2$ & $\gtrsim0.81$ & bl2, bl-BAL, BALQSO\\
 &  & -9934 & $1.73\pm0.18$ & $1.02\pm0.13$ & $1053.5\pm108.9$ & $\gtrsim0.35$ & broad, BALQSO\\
\enddata
\end{deluxetable*}

\startlongtable
\begin{deluxetable*}{cccccccc}
\tablecaption{\civ\ mini-BALs in Keck-HIRES. See \Cref{tab:mBAL_uves} for descriptions of the table contents.\label{tab:mbal_keck}}
\tabletypesize{\scriptsize}
\tablehead{
\colhead{QSO} & \colhead{$z_{em}$} & \colhead{$v$} & \colhead{REW(\lam1548)} & \colhead{REW(\lam1551)} & \colhead{FWHM(\lam1548)} & \colhead{$C_0$} & \colhead{Notes} \\ 
\colhead{} & \colhead{} & \colhead{(\kms)} & \colhead{(\AA)} & \colhead{(\AA)} & \colhead{(\kms)} &\colhead{} & \colhead{}
 }  
\startdata
J100841+362319 & 3.126 & -5728 & $0.18\pm0.03$ & $0.12\pm0.03$ & $321.5\pm56.9$ & $0.14\pm0.02$ & bl1, cmplx\\ 
J101723-204658 & 2.545 & -10666 & $0.03\pm0.00$ & $0.02\pm0.00$ & $101.5\pm33.3$ & $0.12\pm0.05$ & unc \\
J102325+514251 & 3.447 & -4544 & $0.62\pm0.05$ & --- & $288.7\pm35.3$ & $\gtrsim0.45$ & bl1, cmplx\\
  &   & -5931 & $0.17\pm0.01$ & $0.12\pm0.02$ & $292.3\pm40.6$ & $0.12\pm0.01$ & bl1, cmplx\\
J115940-003203 & 2.035 & -4469 & $0.62\pm0.08$ & $0.40\pm0.08$ & $316.9\pm65.1$ & $0.44\pm0.06$ & bl2, cmplx\\
& & -6730 & $0.09\pm0.01$ & $0.08\pm0.01$ & $210.4\pm166.5$ & $0.08\pm0.01$ & bl1, cmplx\\
  &   & -8045 & $0.51\pm0.06$ & $0.37\pm0.07$ & $314.1\pm61.9$ & $0.33\pm0.04$ & cmplx\\
  &   & -9835 & $0.47\pm0.05$ & $0.33\pm0.06$ & $278.2\pm44.5$ & $0.34\pm0.03$ & bl2, cmplx\\ 
J131215+423900 & 2.567 & -5459 & $0.52\pm0.01$ & $0.39\pm0.05$ & $170.2\pm8.2$ & $0.61\pm0.01$ & cmplx, bl2\\ 
 &  & -6433 & $0.85\pm0.02$ & $0.64\pm0.08$ & $211.9\pm10.0$ & $0.78\pm0.01$ & cmplx, bl2-6570, bl2-6872\\ 
&  & -6570 & $0.82\pm0.01$ & $0.61\pm0.06$ & $271.2\pm14.0$ & $0.59\pm0.01$ & cmplx, bl2-6433\\
 &  & -6872 & $0.63\pm0.34$ & --- & $267.0\pm98.5$ & $\gtrsim0.48$ & cmplx, bl2-6433, bl2-7273\\ 
& & -7273 & $0.16\pm0.14$ & --- & $279.7\pm48.2$ & $\gtrsim0.10$ & cmplx, bl2-6872\\ 
&   & -18030 & $0.74\pm0.56$ & $0.39\pm0.35$ & $673.0\pm250.2$ & $\gtrsim0.21$ & broad\\
&  & -19778 & $2.84\pm0.06$ & $2.00\pm0.13$ & $1337.8\pm46.2$ & $\gtrsim0.43$ & broad, bl1, unc\\ 
J141719+413237 & 2.024 & -4136 & $2.37\pm0.03$ & $1.98\pm0.04$ & $512.3\pm15.1$ & $0.89\pm0.01$ & bl2-4483, cmplx, BALQSO\\ 
& & -4483 & $0.92\pm0.05$ & $0.68\pm0.06$ & $196.4\pm20.7$ & $0.93\pm0.02$ & bl2-4136, cmplx, BALQSO \\ 
J160455+381214 & 2.551 & -11104 & $0.49\pm0.01$ & $0.36\pm0.06$ & $269.2\pm9.5$ & $0.36\pm0.01$ & var (Apr 2005), cmplx \\
  &   & -11098 & $0.41\pm0.01$ & $0.30\pm0.05$ & $275.6\pm12.3$ & $0.30\pm0.01$ & var (May 2009), cmplx\\ 
J160843+071508 & 2.877 & -4710 & $0.23\pm0.02$ & $0.14\pm0.02$ & $161.1\pm17.0$ & $0.36\pm0.04$ & var (Apr 2007), cmplx \\
  &   & -4705 & $0.31\pm0.03$ & $0.20\pm0.03$ & $172.8\pm18.9$ & $0.40\pm0.04$ & var (Jul 2008), cmplx \\\ 
J162453+375806 & 3.380 & -23352 & $0.29\pm0.01$ & $0.22\pm0.09$ & $234.7\pm19.0$ & $0.25\pm0.01$ & bl2\\
J175603+574848 & 2.110 & -5274 & $0.34\pm0.05$ & $0.20\pm0.04$ & $141.3\pm19.7$ & $0.65\pm0.09$ & var (Oct 2004), bl2, cmplx\\
  &   & -5276 & $0.43\pm0.03$ & $0.27\pm0.04$ & $153.4\pm12.4$ & $0.64\pm0.04$ & var (Aug 2006), bl2, cmplx\\
  &   & -5274 & $0.41\pm0.02$ & $0.26\pm0.04$ & $160.0\pm11.4$ & $0.61\pm0.03$  & var (Jul 2008), bl2, cmplx\\
J212329-005052 & 2.262 & -8138 & $0.12\pm0.01$ & $0.09\pm0.01$ & $74.3\pm4.9$ & $\gtrsim0.33$ & bl2, cmplx, var (Aug 2006)\\
 &   & -9117 & $0.16\pm0.01$ & $0.11\pm0.01$ & $103.9\pm6.3$ & $0.31\pm0.01$ & cmplx, var (Aug 2006)\\
  &   & -11501 & $0.23\pm0.05$ & $0.12\pm0.04$ & $178.6\pm28.1$ & $0.55\pm0.11$ & cmplx, bl1, unc, var (Aug 2006)\\
  &   & -12337 & $0.30\pm0.01$ & $0.23\pm0.06$ & $197.6\pm9.8$ & $0.30\pm0.01$ & cmplx, bl2, var (Aug 2006)\\
  &   & -13546 & $0.06\pm0.05$ & $0.03\pm0.02$ & $82.3\pm54.2$ & $\gtrsim0.10$ & cmplx, bl2, unc, var (Aug 2006)\\
J231324+003444 & 2.083 & -7490 & $0.18\pm0.07$ & $0.10\pm0.05$ & $169.6\pm57.0$ & $0.34\pm0.14$ & bl2, cmplx\\
  &   & -8245 & $0.32\pm0.02$ & $0.24\pm0.02$ & $233.0\pm20.7$ & $0.27\pm0.01$ & cmplx \\
  &   & -9438 & $0.22\pm0.06$ & $0.11\pm0.05$ & $397.7\pm234.7$ & $\gtrsim0.10$ & cmplx, unc\\
  &    & -11036 & $4.12\pm0.44$ & $2.39\pm0.31$ & $1275.3\pm131.9$ & $\gtrsim0.70$ & broad, cmplx\\ 
  &   & -23025 & $1.31\pm0.14$ & $0.80\pm0.15$ & $587.9\pm76.1$ & $0.56\pm0.05$ & broad\\ 
  &   & -30887 & $0.66\pm0.05$ & $0.60\pm0.07$ & $936.4\pm188.9$ & $0.13\pm0.00$ & mBAL?, broad, unc\\ 
\enddata
\end{deluxetable*}

\subsection{Mini-BAL Properties \& Statistics}

Overall we find 105 high-velocity \civ\ mini-BALs in 44 quasars out of 638 total quasars in our sample. \Cref{tab:basics} provides a more detailed break down of some of the sample statistics. For example, 44 quasars (38 non-BALQSOs) have $\geq1$ \civ\ mini-BAL while 25 quasars (21 non-BALQSOs) have $\geq2$ \civ\ mini-BALs. 

\begin{figure}
\centering
\includegraphics[width=0.45\textwidth]{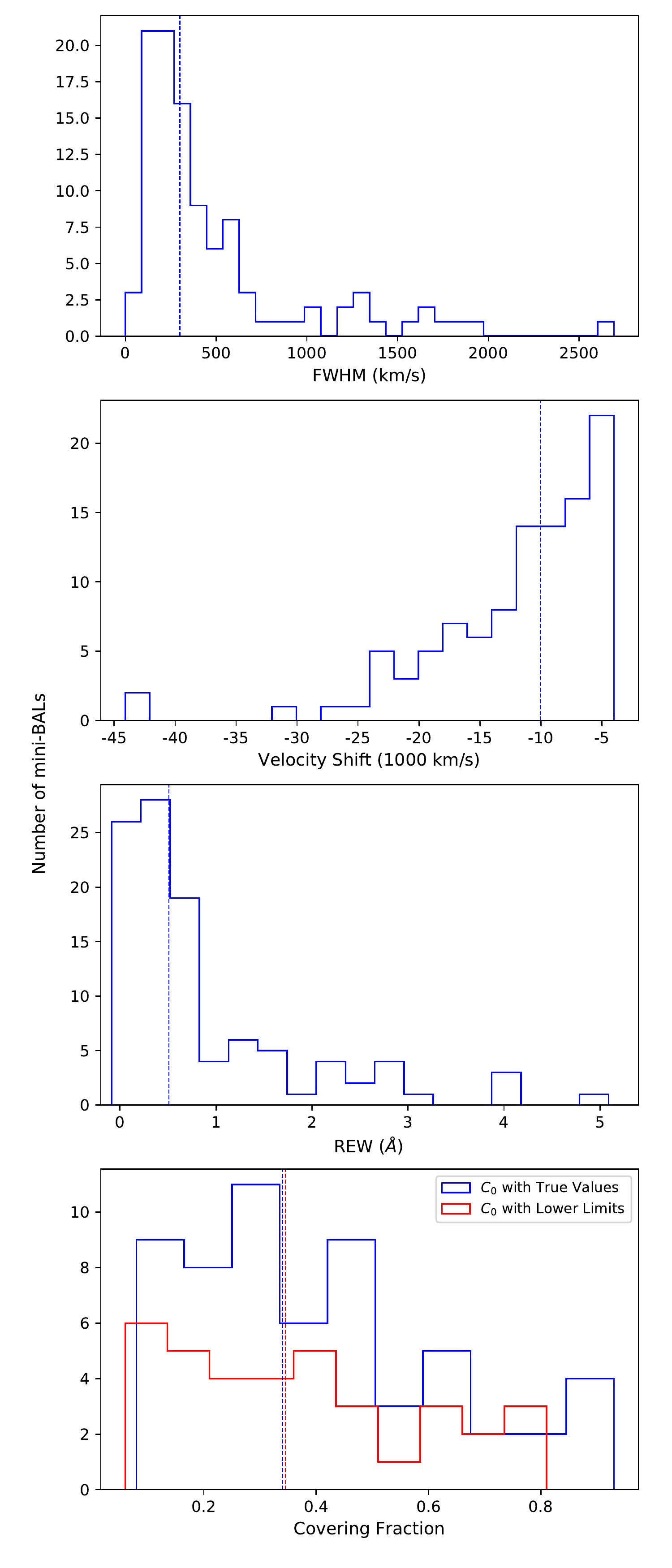}
\caption{\civ\ mini-BAL parameter distributions, from top panel to bottom: FWHM(1548), velocity shift, REW(1548), and covering fraction. Rare cases with large uncertainties (due to poorly constrained fits, Section 3) are excluded. The dashed vertical lines show the median values (300 \kms, $-10.0\times10^3$ \kms, 0.51 \AA, and 0.34, respectively, from top to bottom panel) in the distributions.\label{fig:distribution}}
\end{figure}

\Cref{fig:distribution} shows the numbers of mini-BALs having different values of FWHM(1548), velocity shift, REW(1548), and covering fractions. The top panel in \Cref{fig:distribution} shows that most of the mini-BALs in our study are relatively narrow. In particular, the median FWHM(1548) is 300 \kms\ and $\sim25$\% of the mini-BALs have FWHM(1548) $\lesssim 200$ \kms. Similarly, most of the mini-BALs are weak, with median REW(1548) $\sim$ 0.51 \AA . Thus we find that many of the mini-BALs in our study are weaker and narrower than previous studies, e.g., those based moderate-resolution spectra like the SDSS \citep{Hamann04, Paola08}. Those previous studies set conservatively large FWHM thresholds to avoid broad blends of multiple narrow absorption lines masquerading as mini-BALs. This contamination problem is avoided by the high-resolution spectra used here, such that we can securely identify mini-BALs as outflow lines with smooth BAL-like profiles down to much narrower line widths. The large fraction of narrow mini-BALs in our study also indicates that the large FWHM thresholds used in previous studies exclude the majority of mini-BALs present in quasar spectra. For example, applying the conservatively large threshold FWHM(1548) $>$ 700 \kms\ adopted by \cite{Paola08} for their SDSS study to our sample would miss 75\% of the mini-BALs in our catalog. 

The distribution of velocity shifts shown in \Cref{fig:distribution} rises steeply toward small outflow velocities. Note that the distribution is truncated at v$\sim -4000$ \kms\ by our selection constraint (Section 2.2). Overall, the velocity range of the mini-BALs is similar to BALs, including 3 mini-BALs ($\sim3$\% of the sample) at high velocities $<-30,000$ \kms . These rare high-velocity cases are highlighted by orange boxes in \Cref{fig:mbal_all}. Conversely, $\sim35$\% of the mini-BALs in our sample are at low velocities from $-8000$ to $-4000$ \kms . 

The bottom panel in \Cref{fig:distribution} shows separately the distributions of well-measured covering fractions for which we have direct values (from saturated lines and/or resolve doublets) or firm lower limits (Section 3). These covering fraction distributions indicate that most and possibly all \civ\ mini-BALs in our catalog have partial covering. In particular, all of the well-measured $C_0$ values are $<$1. Therefore, considering only well-measured systems, 100\% have $C_0<1$, $\sim75\%$ have $C_0\lesssim0.5$, and $\sim7\%$ have definite $C_0 < 0.1$. The weak mini-BALs with well-measured $C_0 \lesssim 0.1$ are highlighted by green boxes in \Cref{fig:mbal_all}. 


Finally, we searched the literature for published studies that might suggest the quasars in our sample were targeted for VLT-UVES or Keck-HIRES spectroscopy because of they were known or suspected ahead to have outflow lines. Such targeted observations might bias the mini-BAL statistics in our study, depending on the specific targeting criteria. We find that none in the UVES sample were targeted for outflow lines in the velocity range $v\lesssim -4000$ \kms\ we consider. In the HIRES sample, we find only two quasars, J100841+362319 and J102325+514251, that were targeted (by our team for another study) to have complexes of AALs that could include outflow lines \citep{Chen19, Simon10b, Simon12}. All of the other quasars appear to have been targeted for studies of i) intervening absorption lines (e.g., DLAs and LLSs) that should be unbiased for mini-BALs in the spectra, or ii) associated absorption lines \citep{Odorico04} at low velocities that are not part of our catalog. 

The raw fraction of quasars with at least one mini-BAL in our study, 44/638 $\approx$ 7.0\%, is a lower limit for the line type we consider because systems can be missed due to blends, limitations in the spectral coverage, or bad/noisy spectral regions. Correcting for the incomplete velocity coverage of the spectra (\Cref{fig:n_v}) and conservatively excluding all quasars known to be targeted for any type of intrinsic/associated absorption lines, we estimate that $\sim$9\% of quasars have at least one mini-BAL meeting the definitional requirements of our study (v$\,\lesssim -4000$ \kms\ and $70 \lesssim$ FWHM(1548) $\lesssim 2000$ \kms, Section 2.2). Excluding BAL quasars from the parent sample makes only a small difference, such that $\sim$8\% of non-BAL quasars have a mini-BAL as defined by our study.

\begin{deluxetable}{cc}
\tablecaption{Number of quasars with \civ\ mini-BALs and number of \civ\ mini-BALs in our catalog. Results for all quasars and non-BAL quasars in our dataset are listed separately. \label{tab:basics}
}
\tablehead{
\colhead{Categories} & \# of QSOs (\# of mini-BALs)
}
\startdata
QSOs with $\geq1$ mini-BALs & 44 (105)\\
QSOs with $\geq2$ mini-BALs & 25 (86) \\
non-BALQSOs with $\geq1$ mini-BALs & 38 (91) \\
non-BALQSOs with $\geq2$ mini-BALs & 21 (74) \\
\enddata
\end{deluxetable}

\subsection{Correlation Tests}

\begin{figure}
\centering
\includegraphics[width=0.45\textwidth]{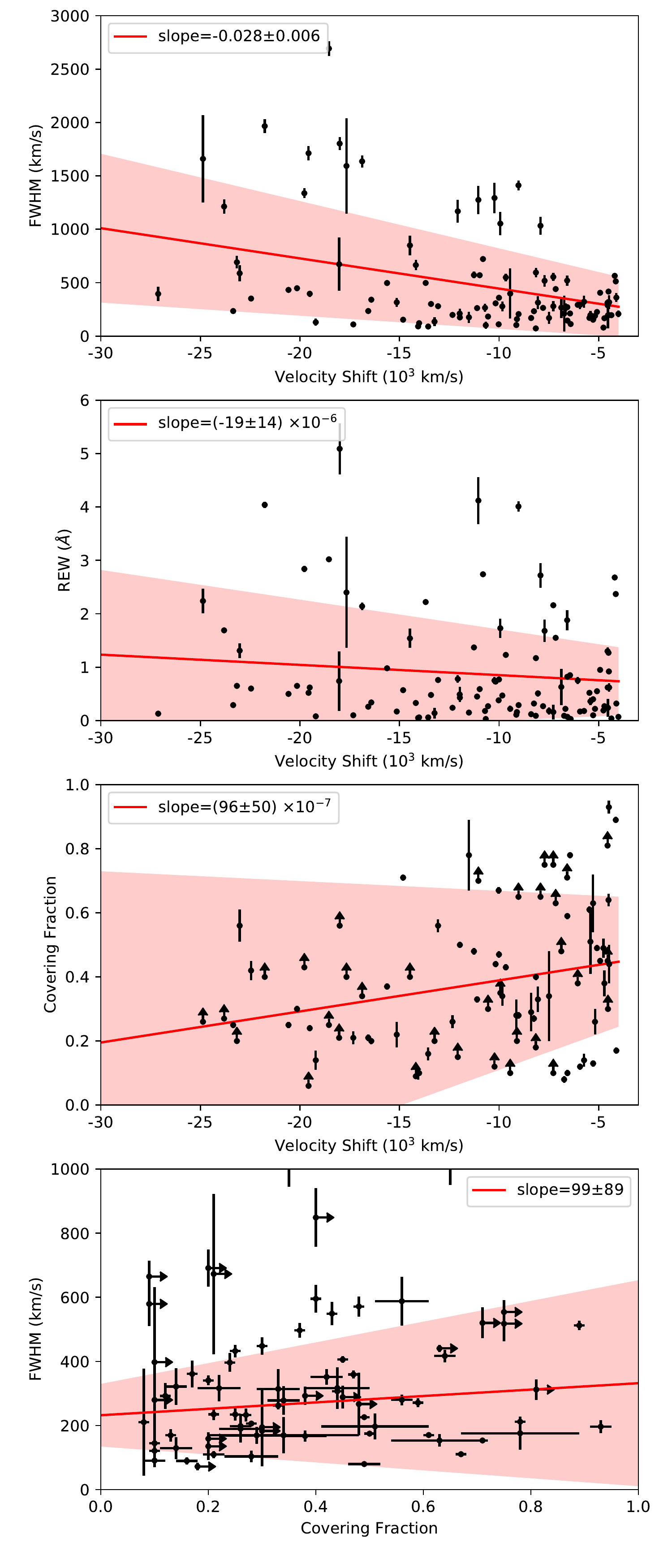}
\caption{Mini-BAL parameter relationships. From top panel to bottom: FWHM(1548) vs. velocity shift; REW(1548) vs. velocity shift; covering fractions $C_0$ vs. velocity shift; and FWHM(1548) vs. covering fraction. As in \Cref{fig:correlation}, rare cases with large uncertainties (estimated to be larger than their formal values returned from the fits) are excluded. The red lines show linear least squares fits to the plotted points. In the two bottom panels, the fits apply only to $C_0$ with derived specific values. The linear slopes and associated $1\sigma$ errors are given in each panel. The shaded regions indicate the $3\sigma$ uncertainties in the linear fits.\label{fig:correlation}}
\end{figure}

\Cref{fig:correlation} shows various \civ\ mini-BAL parameters plotted against the velocity shift and covering fraction to test for trends. There is generally large scatter. The only significant trend is for larger FWHM(1548) at larger velocity shifts (top panel). This relationship has a probability of occurring by chance of $P=0.01\%$ in the Pearson correlation test, which is considered to be extremely statistically significant by conventional criteria. Although there is considerable scatter in the FWHM(1548) values at large velocity shifts, the trends indicates that  typical mini-BAL FWHMs increase by a factor of $\sim$3 from $-$8000 to $-$30,000 \kms . 

There is also tentative trend for mini-BALs at larger velocity shifts to have smaller covering fractions (third panel from top). The probability of this correlation occurring by chance $P=6\%$. The broader lines with smaller covering fractions and nominally shallower troughs result in the lines having no significant trend in REW(1548) with velocity shift (second panel from the top). 

\subsection{Mini-BALs in Outflow Complexes}

\begin{figure*}
\centering
\includegraphics[width=1\textwidth]{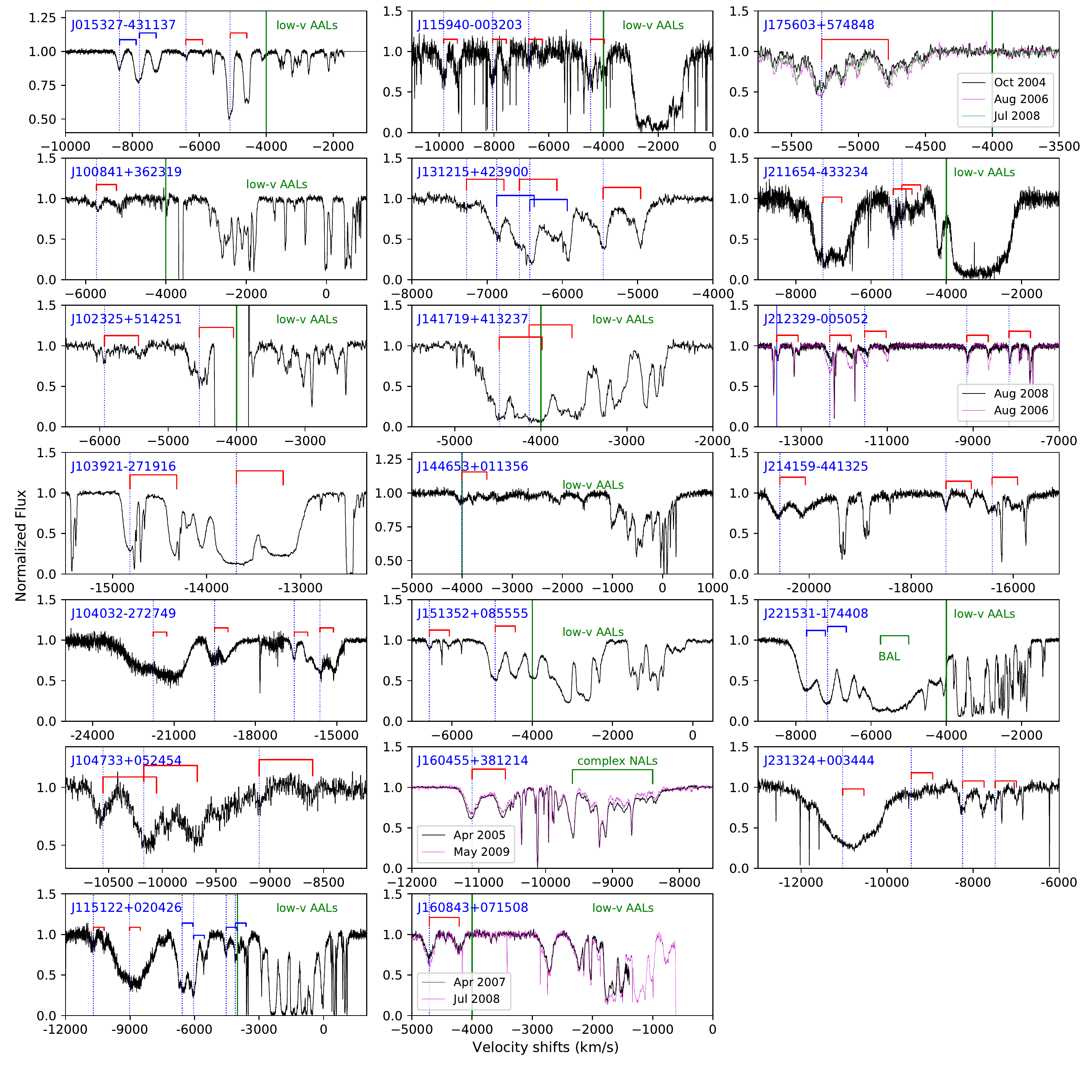}
\caption{Normalized spectra of all \civ\ mini-BALs that appear in multi-component outflow complexes (having 3 or more \civ\ absorption doublets that appear related to each other based on similar profiles and velocity shifts $\lesssim$3000 \kms\ apart, sometimes including low-velocity lines not in our catalog, see also `cmplx' in \Cref{tab:mBAL_uves,tab:mbal_keck}). The vertical green solid lines mark the velocity threshold at $-4000$ \kms\ for consideration in our mini-BAL catalog. The blue dotted lines mark the centroids of mini-BAL profiles in \civ\ \lam 1548, while the red and blue brackets mark \civ\ doublets included in our mini-BAL catalog. Blue brackets indicate mini-BAL pairs that appear to be line-locked (Section 4.5). \civ\ absorption lines not in our catalog (AALs, BALs, or low-velocity systems) are marked by green brackets above the spectra. The velocities on the horizontal axis pertain to the short-wavelength lines in the doublets. Spectra from different observing epochs (when available) are over-plotted in different colors.\label{fig:complex}}
\end{figure*}

\Cref{fig:complex} shows an expanded view of all \civ\ mini-BALs that belong to outflow absorption-line complexes as defined in Section 4.0, e.g., with 3 or more components that appear physically related less than 3000 \kms\ apart (see also \Cref{tab:mBAL_uves,tab:mbal_keck}). These absorption-line complexes are interesting because they trace spatially complex outflow structures near the quasars. Roughly 45\% of the quasars with mini-BALs in our study have an outflow complex by this definition, and $\sim53\%$ of all mini-BALs in our catalog appear in such a complex. The outflow complexes in J100841+362319, J102325+514251 and J212329-005052 were studied in detail by \citep{Chen19, Simon10, Hamann11}. Note that some of the lines in these complexes are narrower and/or at smaller velocity shifts than we consider for the mini-BALs in our study. Their velocity shifts overall span a wide range from $\sim-22,000$ km/s to $\sim0$ \kms\ while the FWHMs in the complex components range from $\sim70$ to $\sim2000$ \kms, i.e., the full range of FWHMs considered in our study. 

\subsection{\pv\ and Other Ions}

\begin{figure*}
\centering
\includegraphics[angle=90,scale=0.5]{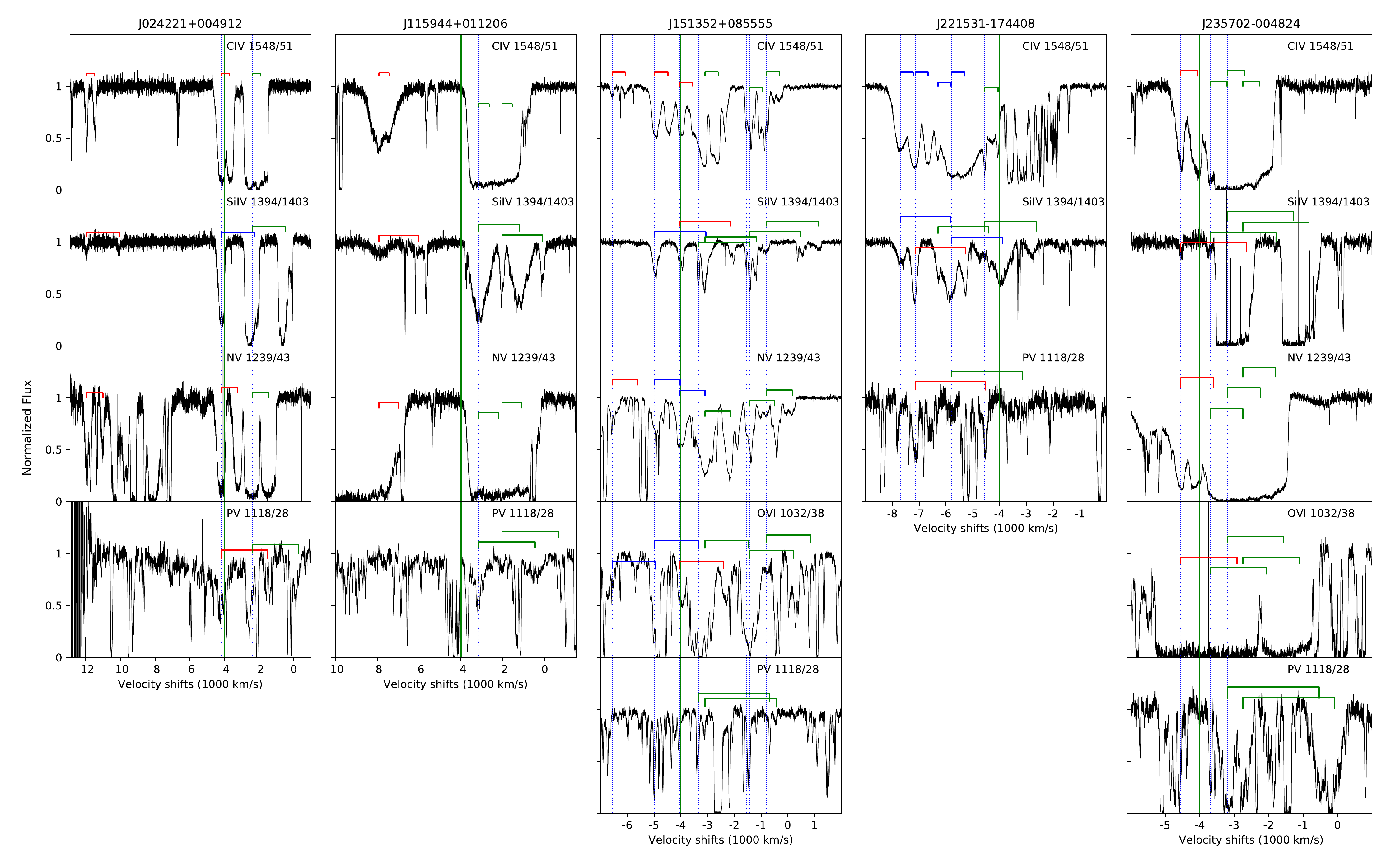}
\caption{Normalized multiple ion absorption profiles in five quasar spectra showing \pv\ \lam 1118,1128 plotted on a velocity scale relative to the quasar redshift. We highlight the systems that appear to show line locking in blue brackets. See \Cref{fig:complex} for additional notes.\label{fig:pv}}
\end{figure*}

For every \civ\ mini-BAL detected in our survey, we search for other absorption lines at the same velocity shift, such as \lya\ \lam 1216, \siiv\ \lam 1393,1403, \nv\ \lam 1239,1243, \ovi\ \lam 1032,1038, \pv\ \lam 1118,1128, \siii\ \lam 1527, \cii\ \lam 1335. None of the low-ionization lines are detected. \pv\ absorption is particularly important as an indicator of large outflow column densities (Sections 1 and 5). Thus we expand our search for \pv\ absorption to include strong outflow systems at lower velocity shifts than our survey limit v$\, < -4000$ \kms . \Cref{fig:pv} shows all of the outflow systems where \pv\ absorption is clearly detected. One of the quasars with \pv\ absorption in \Cref{fig:pv} is a BAL, J221531-174408. The others are mini-BALs or some type of broad outflow AALs. To our knowledge, there is only one other published report of \pv\ absorption in an individual non-BAL quasar \citep[by][]{Hamann19b}. However, the study of median quasar spectra by \cite{Hamann19a} shows that BAL \textit{and} mini-BAL systems typically have \pv\ absorption at a fraction of the strength of \civ . The actual number of quasars with \pv\ absorption in our mini-BAL sample is likely to be larger what is shown in \Cref{fig:pv} because secure detections of \pv\ lines can be thwarted in our study by 1) unrelated absorption lines in the \lya\ forest, 2) observed wavelength coverages that usually do not include \pv\, and 3) the spectra often having low signal-to-noise ratios at short observed wavelengths where the \pv\ lines are found. 

All of the mini-BAL systems with securely detected \pv\ lines have saturated absorption in \civ, \siiv\ and \ovi\ (when available) based on $\sim$1:1 doublet ratios or deep flat-bottomed troughs that reach near-zero intensities. This confirms expectations from photoionization models that measurable \pv\ lines should be accompanied by large optical depths in these other lines, specifically $\tau \gtrsim 1000$ in \civ . The measured \pv\ absorption troughs in \Cref{fig:pv} also tend to be (much) narrower than \civ\ and other high-ionization lines like \ovi . These results are in good agreement with other measurements of \pv\ in individual BAL quasars \citep[][and refs therein]{Chamberlain15, Capellupo17, Moravec17} and with the median composite spectra of large BAL and mini-BAL quasar samples presented in \citep{Hamann19a}. 

\begin{figure*}
\centering
\includegraphics[angle=90,scale=0.5]{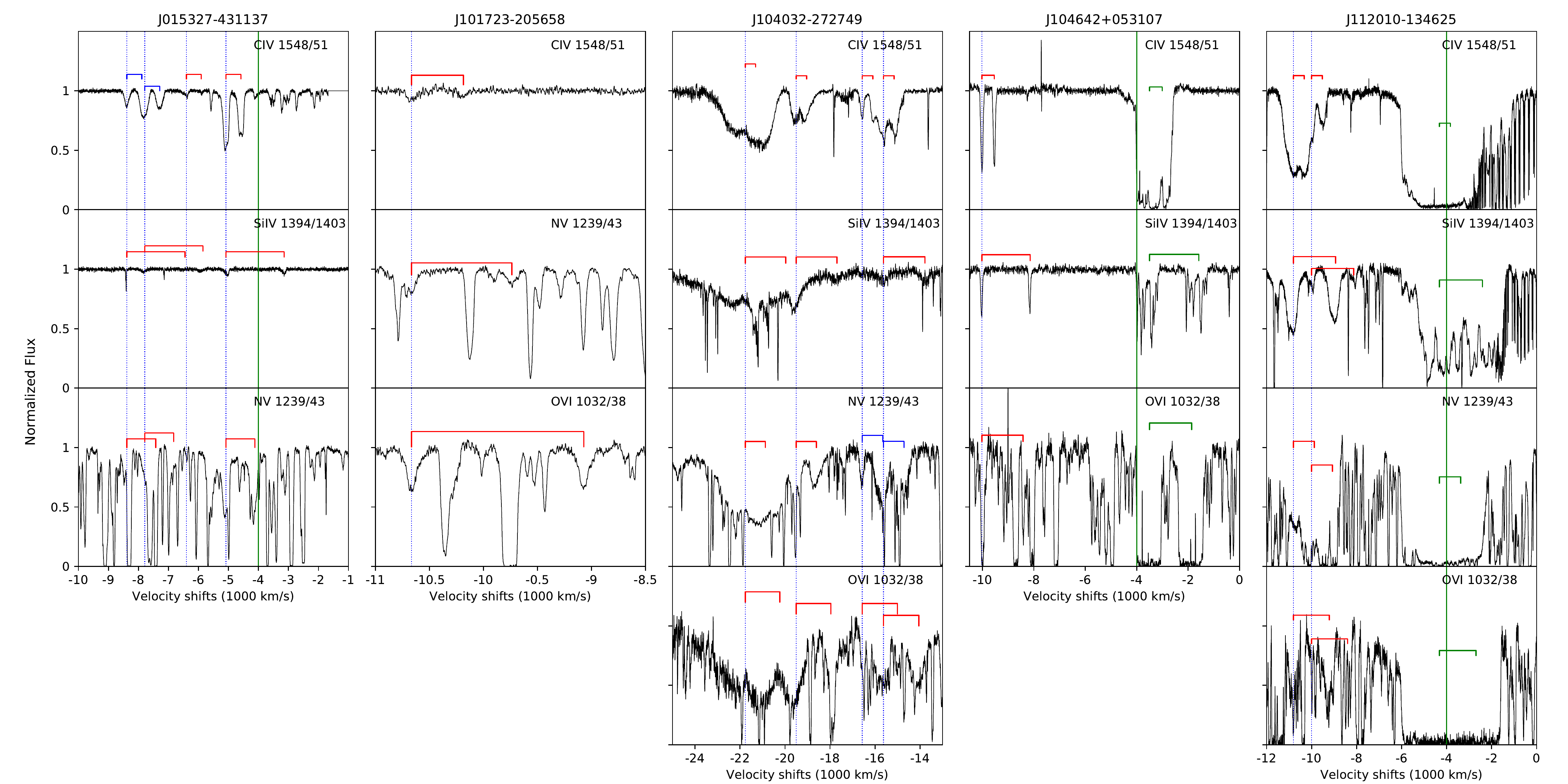}
\caption{Normalized multi-ion absorption profiles on a velocity scale relative to the quasar redshift. See \Cref{fig:pv} for additional notes.\label{fig:multi_ions}}
\end{figure*}

\begin{figure*}
\centering
\setcounter{figure}{7}
\includegraphics[angle=90,scale=0.5]{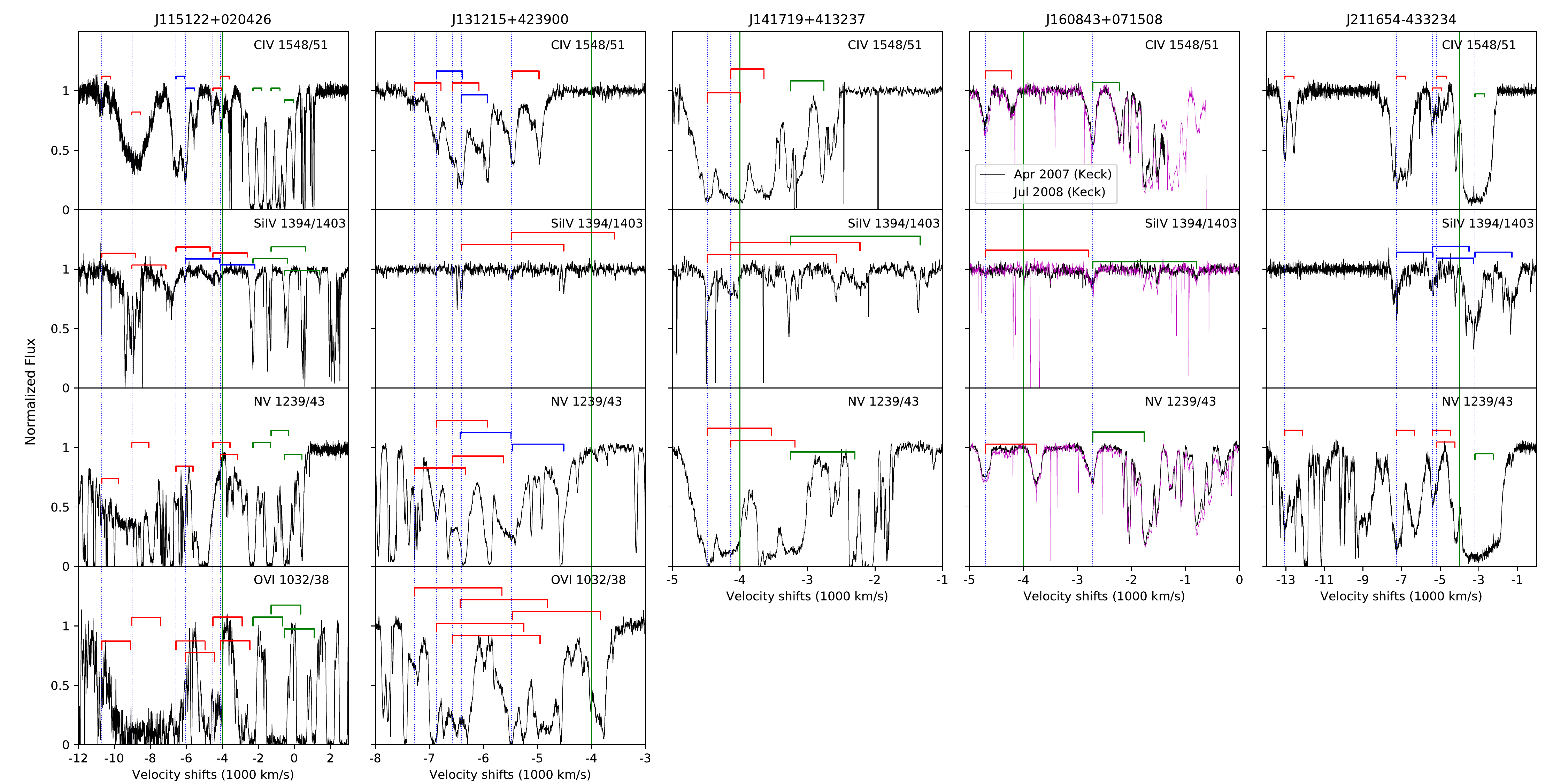}
\caption{\textit{continued.}}
\end{figure*}

\Cref{fig:multi_ions,fig:J160455} show all of the remaining mini-BAL systems in our catalog with lines other than \civ\ clearly detected. All of these multi-ion systems, including those with \pv\ detections in \Cref{fig:pv}, exhibit a general trend for deeper absorption troughs and wider profiles in strong transitions of higher ions such as \civ, \nv, and \ovi\ compared lower ions like \siiv\ and weaker lines due to low abundance like \pv . The mini-BAL systems with resolved doublets in \Cref{fig:pv,fig:multi_ions} indicate that all of the lines are saturated with moderate depths due to partial covering. This implies that the progression in line depths and velocity widths with ionization and line optical depth is directly tied to a trend in the covering fractions, namely, that the strong lines of higher ions form in spatially larger regions while weaker transitions and lines of lower ions tend to form in smaller pockets of higher column density gas. These results are consistent with other  studies of individual outflow quasars and composite BAL and mini-BAL spectra (e.g., citations in previous paragraph). We discuss the implications of these results for the physical conditions and spatial structure of quasar outflows in Section 5.

\subsection{Line-locked mini-BALs}

Line-locked absorption-line systems have distinct components at velocity separations equal to prominent doublets such as the \civ, \nv, or \siiv . Lines formed in different outflow clumps can become locked at these doublet separations due shadowing effects in radiatively-driven outflows. Line-locked systems can therefore be a signature of outflows driven by radiative forces in the vicinity of a quasar \citep{Milne26, Scargle73, Braun89}. It is also possible that lines appearing at the doublet separation in observed spectra result chance alignments of physically unrelated absorption-line clouds. However, observations of multiple line-locks in the same spectrum argue strongly for the reality of physical line locks in at least some cases \citep[e.g.,][]{Ganguly08, Hamann11}. The large statistical study by \citet{Bowler14} showed that roughly two-thirds of SDSS quasars with multiple \civ\ NALs at speeds up to $\sim12,000$ km/s have at least one line-lock pair. Those results indicate that physical line-locking due to radiative forces is both real and common in quasar outflows.

In our study, five pairs of mini-BALs in four quasars appear line-locked at the \civ\ doublet separation (498 \kms ) to an accuracy $<$20\% of the FWHM of the narrowest fitted line in the pair. This amounts to 4/25 = $\sim$16\% of quasars with multiple \civ\ mini-BALs having at least one line-locked \civ\ pair. These five line-lock pairs are marked by in blue brackets in \Cref{fig:complex}. We also find apparent line-locks in the doublet lines of other ions, \ovi, \nv\ and \siiv . These line-lock cases are marked by blue brackets in \Cref{fig:pv,fig:multi_ions}. The velocity offsets in these cases match the doublet separations very well, within $\lesssim10$\% of the line FWHMs. Including all of these systems in the tally above, we find that 7/25 = 28\% of quasars with multiple mini-BALs have at least one line-lock pair. 

\subsection{Variable mini-BALs}

A small subset of the quasars in our study were observed more than once with VLT-UVES or Keck-HIRES. We check for line variability between these observations by visual inspecting the normalized spectra plotted on top of each other (e.g., as in \Cref{fig:complex,fig:pv,fig:multi_ions}). We specifically search for flux mismatches in absorption troughs accompanied by good matches in the continuum adjacent to the troughs. The main uncertainty in these assessments is the continuum placement (not signal-to-noise ratios in the spectra). We record only obvious occurrences of line variability with `var' in the Notes in \Cref{tab:mBAL_uves,tab:mbal_keck}. We then fit the lines separately in each observing epoch for these definite variability cases. 

\begin{figure}
\centering
\includegraphics[width=0.45\textwidth]{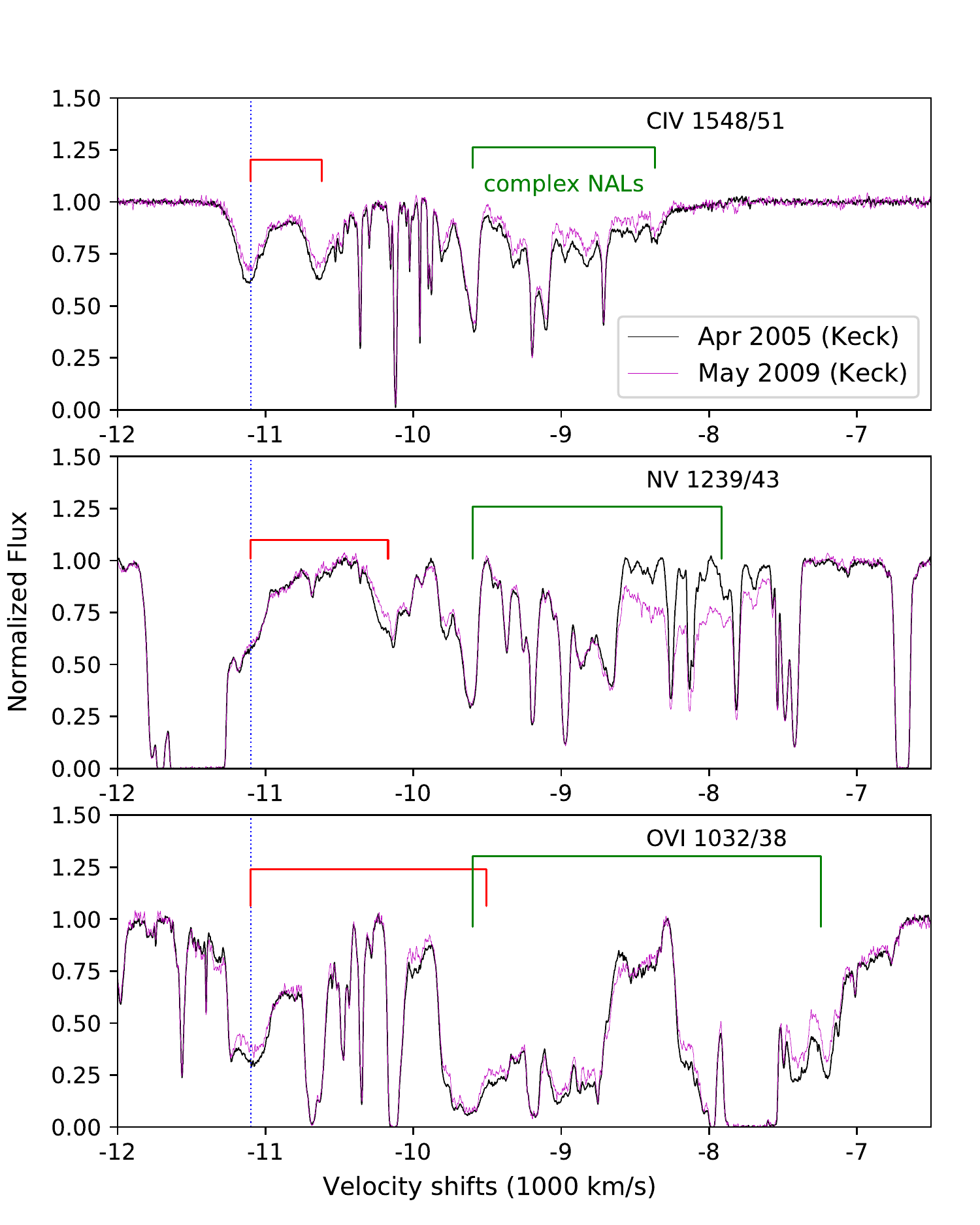}
\caption{Normalized multi-ion absorption profiles in J160455+381214 on a velocity scale relative to the quasar redshift. Variability for multiple ion absorptions in two observations. There are obvious changes in the absorption-line strength between the two epochs. See \Cref{fig:pv} for additional notes.\label{fig:J160455}}
\end{figure}

We identify four quasars in our sample with definite mini-BAL variability, all with changes in more that one mini-BAL. They are J160455+381214 shown in \Cref{fig:J160455}, J160843+071508 in \Cref{fig:multi_ions}, and J175603+574848 and J212329-005052 in \Cref{fig:complex}. The variability in J160455+381214 was studied previously by \citet{Misawa05, Bachev05, Misawa07c, Carini07, Misawa14, Horiuchi16} using data from other telescopes, mainly Subaru. The variability in J160843+071508 was studied by \citet{MacLeod12, Chen15} using spectra from the SDSS, and the variability in J212329-005052 was studied in detail by \citet{Hamann11} using the same data as our current study. There is no previous study to our knowledge of the line variability in J175603+574848. 

We briefly summarize the measured mini-BAL variabilities as follows. The fitted \civ\ mini-BALs in J160455+381214 changed by $\sim$20\% in REW caused by a $\sim$20\% change in covering fraction with no obvious changes ($\lesssim$2\%) in FWHMs or velocity shifts over 4.07 yr in the observer frame (1.15 yr in the quasar frame). Changes are also evident in \ovi\ and with greater magnitude in \nv\ (\Cref{fig:J160455}). The \civ\ mini-BAL included in our catalog for J160843+071508 changed by $\sim35\%$ in REW, $\sim11\%$ in covering fraction, $\sim7\%$ in FWHM, with no obvious change in velocity shifts in 1.25 yr (0.32 yr in the quasar rest frame). This quasar also shows weak variability in \nv\ and \siiv\ absorptions (\Cref{fig:multi_ions}). The lone \civ\ mini-BAL we record for J175603+574848 (in a complex blend) changed by $\sim24\%$ in REW, $\sim6\%$ in covering fraction, $\sim13\%$ in FWHMs, and no changes in velocity shifts within 3.76 yr (1.21 yr in the quasar rest frame). The five \civ\ mini-BALs in J212329-005052 show the changes by $\sim0-\sim56\%$ in the REWs, $\sim25\%$ in the covering fractions, and no significant changes in the FWHMs or velocity shifts in 2.02 yr (0.62 yr in the quasar rest frame). This quasar shows similar variabilities in its \ovi\ absorption lines.

\section{Summary \& Discussion}

We present a unique new catalog of high-velocity \civ\ mini-BALs identified by visual inspection of 638 quasars in the VLT-UVES and Keck-HIRES archives (\Cref{tab:mBAL_uves,tab:mbal_keck}). One important feature of our mini-BAL catalog is that the high-resolution and high-signal-to-noise spectra in these archives are much more sensitive to weak and narrow mini-BALs than all previous outflow line surveys using medium-resolution spectra such as the SDSS. We identify mini-BALs based on smooth rounded BAL-like absorption profiles with velocity widths in the range $70 \lesssim$ FWHM(1548) $\lesssim 2000$ \kms . The approximate upper bound on FWHM avoids BALs while the lower bound limits contamination from unrelated intervening absorption lines (see Section 2.2). We also limit our mini-BAL catalog to velocity shifts $\lesssim -4000$ \kms\ from the quasar frame to avoid associate absorption lines that can reside complex blends or have square profiles that are not BAL-like and unlikely to form in BAL-like outflows. We then fit every \civ\ mini-BAL meeting these criteria to obtain basic line properties like REW(1548) (which ranges from 0.03 to 5.09 \AA ), FWHM(1548) (from 70 to 2693 km/s), and the line-of-sight covering fraction, $C_0$ (from $\sim$0.06 to $\sim$1). 

We estimate that roughly $9\%$ of quasars in our study have at least one \civ\ mini-BAL meeting the selection criteria described above, after correcting for limited spectra coverage and noisy spectral regions but not accounting for undercounting due to blends (Section 4.1). This fraction is similar to the value quoted by \citet{Paola08, Hamann12} who found that \civ\ mini-BALs appear in $\sim11\%$ of bright SDSS quasars. However, there is little overlap in the types of mini-BALs counted in these two studies because of the very different sensitivities, spectral resolutions, and definitional properties of the mini-BALs considered. For example, the SDSS spectral resolution is $\sim$150 \kms\ compared to $\sim$3 to $\sim$8 \kms\ in our study. The SDSS-based study by \cite{Paola08} considered any blueshifted velocity $<$0 \kms\ but also set a very high minimum width threshold at FWHM(1548) $\gtrsim$ 700 \kms\ (to avoid contamination by unresolved blends). In contrast, we count mini-BALs up to 10 times narrower but only if they are at relatively large velocity shifts $v < -4000$ \kms . The steeply-rising numbers of mini-BALs at low velocities in our study (\Cref{fig:distribution}) indicates that we exclude substantial numbers of mini-BALs at smaller velocity shifts. Similarly, the FWHM distribution in our study (\Cref{fig:distribution}) indicates that the SDSS studies substantially undercount mini-BALs by considering only broad ones. Therefore, the true total fraction of quasars with any type of mini-BAL with FWHM(1548) $>$ 70 \kms\ at any blueshifted velocities is probably close to the sum of the two fractions from these studies, e.g., 15\% to 20\%. Furthermore, we need to note that there is an important selection effect in our sample. Most of the parent sample of spectra for our study were taken for intervening absorption lines. Therefore, it is very probable that many objects showing intrinsic absorptions were deliberately not chosen for Echelle spectroscopy, as the intrinsic absorbers complicate the studies of intervening absorption. This selection effect will cause another possible underestimation of the frequency of mini-BAL and BALs.

All of the mini-BAL absorbers identified by \civ\ in our catalog are highly ionized based on the absence of accompanying low-ionization lines such as \siii\ or \cii\ and the strong presence of high-ionization lines such as \nv\ and \ovi\ (when within our spectral coverage). \pv\ absorption is detected unambiguously in 2 mini-BAL systems in our catalog, plus three other quasars with broad outflow lines not in the catalog because their velocity shifts are too low (Section 4.4 and \Cref{fig:pv}). These \pv\ detections indicate that the absorption lines of more abundant ions like \siiv, \civ, \nv, and \ovi\ are highly saturated and that the total outflow column densities are very large (at least at the velocities where \pv\ is detected). Photoionization models presented in other studies show, specifically, that \pv\ detections indicate \civ\ optical depths of several hundred to $\sim$1200 and that total column densities $N_H > 10^{22}$ \cmN\ \citep[for solar abundances, with specific column densities depending on the observed line strengths and widths,][]{Hamann98, Leighly09, Leighly11, Capellupo14, Chamberlain15, Moravec17, Hamann19a, Hamann19b}. An intense radiation field with large ionization parameters is needed to produce these large column densities of ionized gas for Pv absorption. \citet{Hamann19a} combined this result with minimum gas densities inferred from excited-state absorption lines (e.g., in median spectra of BAL and mini-BAL quasars) to conclude that the outflows reside typically within a few tens of pc of the central quasars \citep[see also][]{Hamann19b}.


The majority of \civ\ mini-BALs with resolved doublets in our study, and all of the cases with flat-bottomed troughs or accompanying \pv\ detections, indicate that the lines are optically thick with partial covering of the background light source \citep[see also][for similar mini-BAL results]{Moravec17}. Other cases with line blending yield only lower limits on the covering fractions and optical depths (see \Cref{tab:mBAL_uves,tab:mbal_keck}). Altogether, more than half of the \civ\ mini-BALs in our survey overall are confirmed to be optically thick with partial covering. This requires typical absorbing structures that are not much larger than the projected area of the quasar emission regions. Given that our catalog considers only large velocity shifts ($v<-4000$ \kms ) from the \civ\ broad emission lines, most or all of the mini-BAL partial covering applies to the quasar continuum source. This is the UV-emitting accretion disk $<$0.01 pc across for typical luminous quasars like our sample \citep[see][for a recent discussion]{Hamann19b}. The covering fractions measured in our study, as small as 0.06 in \civ\ (J222006-280323, \Cref{fig:mbal_all}) and 0.03 in \siiv\ (J015327-431137, \Cref{fig:multi_ions}), imply that the outflow absorbing structures can be $\lesssim$0.002 pc across. 

When strong transitions in the higher ions \ovi\ and \nv\ are within our spectral coverage, they tend have deeper and broader absorption troughs compared to \civ . Similarly, the \civ\ lines tend to be deeper and broader than lower-ionization and/or low-abundance lines such as \siiv\ and \pv\ when they are within the wavelength coverage (\Cref{fig:pv,fig:multi_ions}). We infer from this that 1) all of the lines are typically optically thick, and 2) the progression in the line depths and velocity widths is related to their optical depth-dependent covering fractions in spatially inhomogeneous absorbing regions. 

Spatially inhomogeneous (e.g., clumpy) absorbers present a range of column densities and line optical depths across the projected area of the continuum source. Schematic illustrations of this type of absorbing geometry can be found in \cite{Hamann01} and \cite{Hamann04}. This can naturally lead to different derived covering fractions in different lines if they are optically thick over different projected areas. Weak lines with narrower profiles like the \pv\ and \siiv\ mini-BALs form primarily in smaller pockets of high-column density (and perhaps lower-ionization) gas, while the broader and deeper absorption troughs of higher ions like \civ\ and \ovi\ form in lower column density regions that cover larger spatial areas and a wider range of velocities \citep[see also][for more discussion]{Arav05, Arav08, Moravec17, Leighly18, Hamann19a, Hamann19b, Choi20}.

One interesting consequence of this clumpy absorbing geometry is that, without confinement by an external pressure, structures $<$0.002 pc to $<$0.01 pc across have very short survival times without external confinement. The dissipation time for an isolated cloud with only thermal internal velocities is roughly the cloud crossing time at the thermal speed. For a conservatively low temperature of $10^4$ K corresponding to thermal speed (for hydrogen) of $v_{th}\sim 13$ \kms, the dissipation times are $\lesssim$150 yr to $\lesssim$750 yr for the maximum cloud sizes given above. However, the mini-BAL profile widths suggest that the actual internal velocities are much larger. The median mini-BAL width in our study, FWHM $\approx 300$ \kms\ (\Cref{fig:distribution}) indicates that the actual cloud dissipation times are only $\lesssim$6 yr to $\lesssim$30 yr. These short survival times suggest that either 1) the clouds are very near the quasars where the flow times (at the observed outflows speeds, nominally $\sim$10,000 \kms ) are shorter than the survival times, or 2) they created in situ potentially at much larger distances \citep[see][for more discussion]{Chen18, Chen19}. 

The subset of multi-component \civ\ absorption line complexes are an interesting subset of mini-BALs in our catalog because they identify multiple distinct yet physically related absorbing structures in the outflows. Their velocity shifts in our sample range from $\sim-22,000$ \kms\ to $\sim0$ \kms\ and their FWHMs range from $\sim70$ to $\sim2000$ \kms . The lines at higher velocities in these complexes tend to be broader and weaker, with smaller covering fractions, than the lines in the same complex at lower velocities \citep[see also][]{Simon10b, Chen19}. The complexes with such tiny sizes and short survival times indicate they are some type of highly-structured outflows very close to the central quasars. 

Finally, we note that in large surveys $\sim$10-15\% of optically-selected quasars have BALs. The true fraction of quasars with BALs is probably closer to $\sim$20\% after correcting for the selection bias against BALQSOs in these surveys (e.g., \citealt{Hewett03}, \citealt{Trump06}, \citealt{Knigge08}, \citealt{Gibson09}, also \citealt{Hamann12} for a review). This is very similar to our estimate that roughly 15\% to 20\% of quasars with mini-BALs. Observationally, there is no boundary in line widths or velocity shifts separating mini-BALs from BALs. In our study, although we exclude quasars with very broad BALs (because they limit our ability to find mini-BALs), we find that $\sim35\%$ of the BAL quasars that remain have mini-BALs, and in roughly half of those appear in the wings of the BAL toughs, suggesting that the BAL and mini-BAL outflows are physically related. The partial covering situation that is common to these features \citep[see also][]{Hamann19a} suggests that the narrower and generally weaker mini-BALs form in outflow clumps or filaments that cover smaller spatial areas (to produce smaller covering fractions) and smaller ranges in velocity (to produce narrower profiles). BALs might form be spatially larger clumps or in collections of many mini-BAL-like clumps that blend together to produce broader deeper troughs in observed spectra \citep{Hall07, Hamann08, Hamann13}.

Similarly, the narrow mini-BALs in our study, with FWHMs as small as $\sim$74 \kms , have widths overlapping with the narrow absorption lines (NALs) cataloged in other surveys \citep[e.g.][]{Wild08, Nestor08}. There is, again, no clear classification boundary between mini-BALs and outflow NALs. Moreover, some of the mini-BALs in our study reside in multi-component absorption-line complexes that include even narrower lines than our survey threshold FWHM $\gtrsim$ 70 \kms . These results indicate that at least some mini-BALs are physically related to the outflow NALs. More work is needed to understand the physical nature and relationships of all of these different outflow absorption-line features. 

\section*{acknowledgments}

We thank the anonymous referee for helpful comments and suggestions. CC and BM acknowledge financial support from Guangdong Major Project of Basic and Applied Basic Research and Fundamental Research Funds for the Central Universities (Sun Yat-sen University, 20lgpy169). This work was supported in part by the National Natural Science Foundation of China (12073092). CC and FH acknowledge the support of funds from University of California, Riverside, USA and by grant AST-1009628 from the USA National Science Foundation. MM acknowledges the support of the Australian Research Council for \textsl{Discovery Projects} grants DP0877998, DP110100866, and DP130100568.

\bibliography{reference}
\bibliographystyle{aasjournal}

\appendix

\section{Notes on Individual Quasars}

\begin{figure}
\centering
\includegraphics[width=0.45\textwidth]{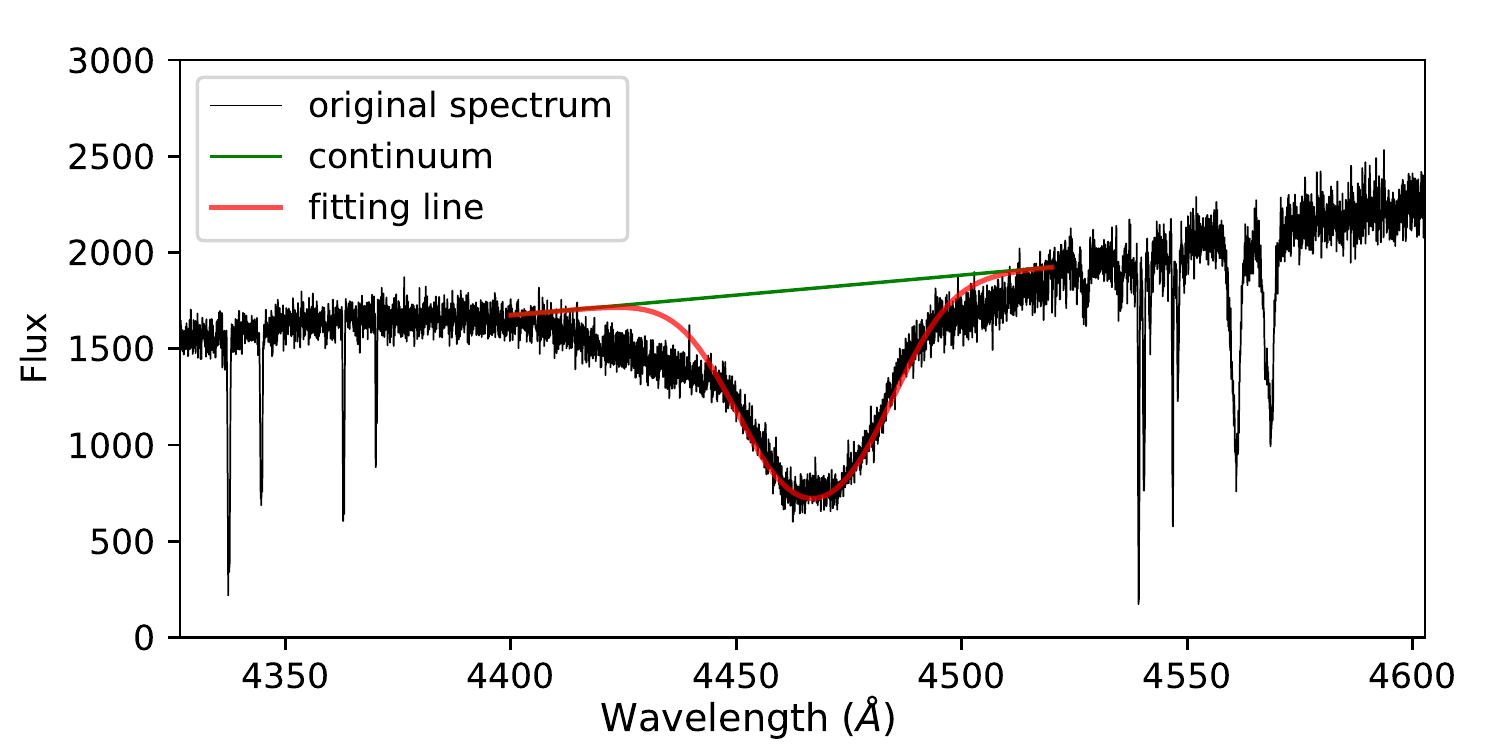}
\caption{We draw a continuum over the mini-BAL system at -18,003 km/s in J024221+004912, and the fitting line is shown in red.\label{fig:J024221_2}}
\end{figure}

This Appendix provides additional notes on mini-BAL systems in our catalog that have ambiguities in the mini-BAL identifications, complex line blends, or uncertainties in the line profile fits that are not already explained in the Notes in \Cref{tab:mBAL_uves,tab:mbal_keck}.
J024221+004912: For the broad system with velocity shift -18,003 km/s (\Cref{fig:J024221_2}), it could be combined by a broad and shallow profile and a narrower and steep profile. But the continuum is uncertain due to the broad feature. Therefore, we only fit the steep and narrower profile. 

\begin{figure}
\centering
\includegraphics[width=0.45\textwidth]{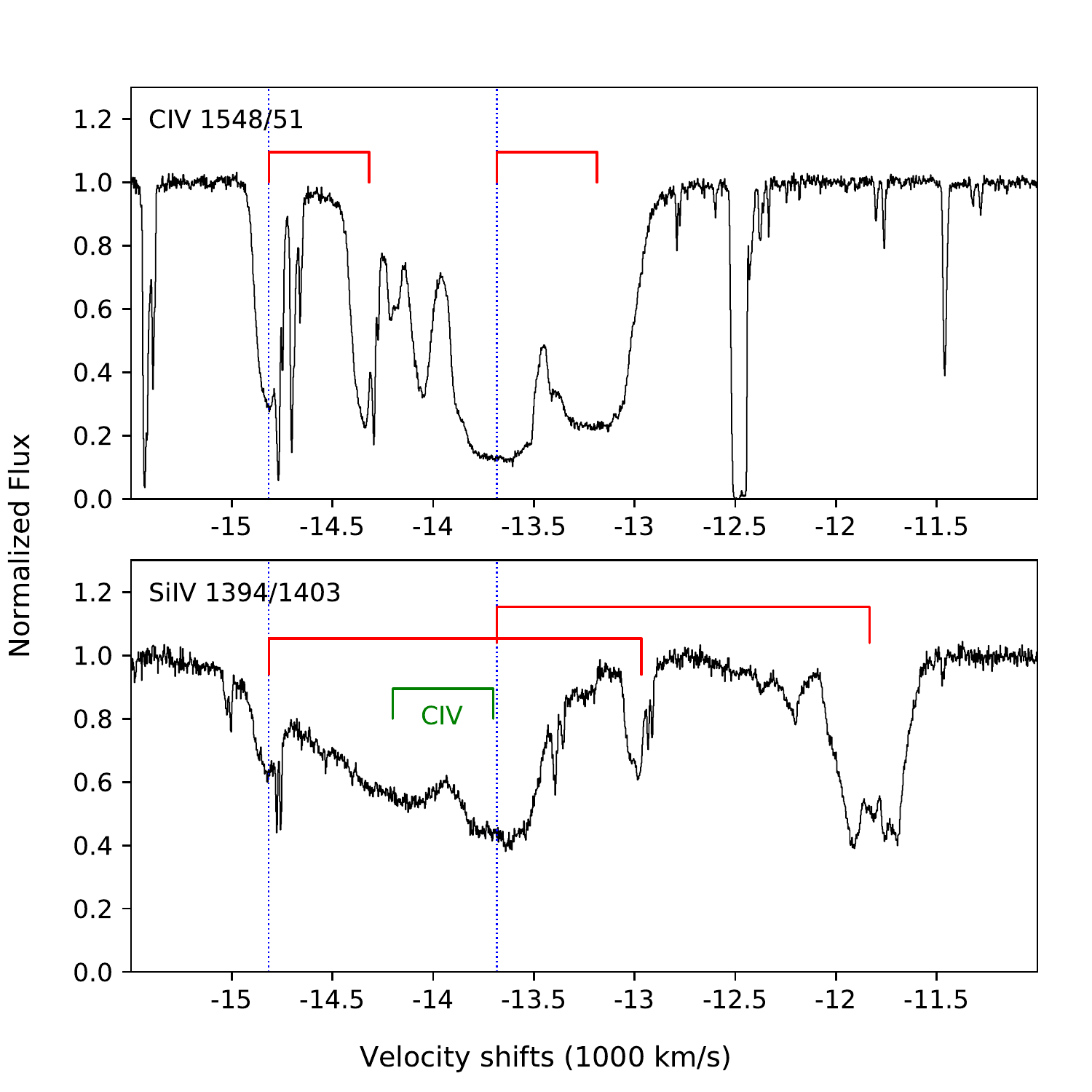}
\caption{J103921-271916: The \civ\ mini-BAL at -42,626 km/s is blended with \siiv\ at -14,816 and -13,684 km/s. See \Cref{fig:mbal_all} for additional notes.\label{fig:J103921}}
\end{figure}

J103921-271916: The \civ\ mini-BAL at -42,626 km/s is blended with \siiv\ (\Cref{fig:J103921}). We estimate the upper and lower limits.

J104032-272749: we confirm the broad system at -21,767 km/s is a \civ\ mini-BAL, because we find other lines \nv\ $\lambda$ 1239, 1243, \siiv\ $\lambda$ 1394, 1403, and \ovi\ $\lambda$ 1032, 1038 at the same redshift with similar but broader blended profile \Cref{fig:multi_ions}. This system is an indistinguishable blend, so we fit the whole feature together.



J160455+381214: this quasar has AAL complexes at $\sim-10000$ to $\sim-8500$ km/s (\Cref{fig:complex}). They are intrinsic because they show partial covering and variability. However, they are too many narrow systems, and we cannot measure all of them. So we make a note of a complex in \Cref{tab:mbal_keck}.


J214225-442018: The system at -13,239 km/s is probably not intrinsic but an intervening system, because it shows neither partial covering nor high ionized absorptions like \nv\ at the same redshift. It is likely a blend.

J231324+003444: The system at -30,887 km/s shows a broad, non-smooth, and asymmetric profile. It could be a blend (\Cref{fig:mbal_all}).

\begin{figure}
\centering
\includegraphics[width=0.45\textwidth]{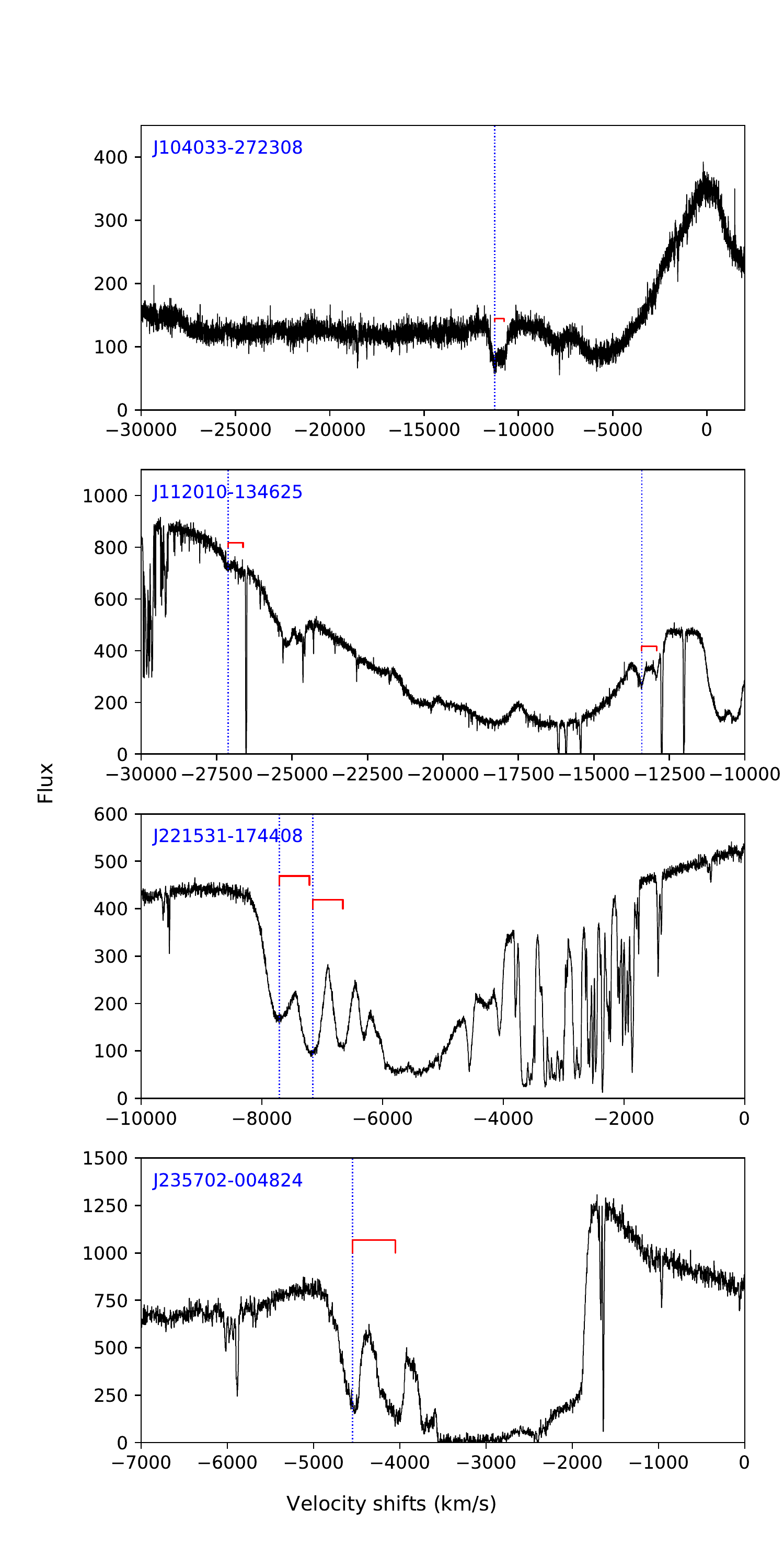}
\caption{Mini-BAL profiles on the wings of BALs or blended with BALs plotted on a velocity scale relative to the quasar redshifts. See \Cref{fig:mbal_all} for additional notes.\label{fig:bl_BAL}}
\end{figure}

The mini-BAL system at -11,259 km/s in J104033-272308, the systems at -13,416 km/s and -27,118 km/s in J112010-134625, the systems at -7155 km/s and -7711 km/s in J221531-174408, and the system at -4549 km/s in J235702-004824, shown in \Cref{fig:bl_BAL}, are on the wings of a BAL or blended with a BAL. So their fits may not good.

\end{document}